\documentclass[11pt]{article}
\usepackage{amsfonts}
\usepackage{amsmath,amssymb}
\usepackage{bbm}
\usepackage{bm}
\usepackage{color}
\usepackage{slashed}
\usepackage{cite}
\usepackage[utf8]{inputenc}
\usepackage[toc,page]{appendix}
\usepackage{graphicx}
\usepackage{placeins}
\usepackage{caption}
\usepackage{subfigure}
\usepackage{booktabs}
\usepackage{comment}
\usepackage{physics}
\usepackage{float}
\usepackage{bbold}
\DeclareGraphicsExtensions{.pdf,.png,.jpg}

\usepackage[colorlinks=false,citecolor=cyan,urlcolor=blue,bookmarks=true,bookmarks=true,bookmarksopen=true,bookmarksnumbered=true,bookmarksopenlevel=3]{hyperref}

\definecolor{airforceblue}{rgb}{0.36, 0.54, 0.66}
\definecolor{steelblue}{rgb}{0.27, 0.51, 0.71}
\definecolor{amber}{rgb}{1.0, 0.49, 0.0}
\definecolor{darkgreen}{rgb}{0.0, 0.5, 0.0}
\definecolor{amber}{rgb}{1.0, 0.49, 0.0}

\graphicspath{{draft/Figures/}}

\allowdisplaybreaks[1]
\def\simg{{\ \lower-1.2pt\vbox{\hbox{\rlap{$>$}\lower6pt\vbox{\hbox{$\sim$}}}}\ }}
\def\siml{{\ \lower-1.2pt\vbox{\hbox{\rlap{$<$}\lower6pt\vbox{\hbox{$\sim$}}}}\ }}

\makeatletter \@addtoreset{equation}{section} \makeatother

\newcommand*{\diff}[1]{\text{d}#1}
\newcommand*{\pder}[2]{\frac{\partial #1}{\partial #2}}
\newcommand*{\der}[2]{\frac{d #1}{d #2}}

\begin{document}

\flushbottom

\begin{titlepage}

\begin{centering}

\vfill

{\Large{\bf
Unitarity in KK-graviton production}}\\\vspace{0.2cm}
{\large {\bf A case study in warped extra-dimensions\\}
}

\vspace{0.8cm}

A.~de~Giorgi$^{\rm a\,}$\footnote{arturo@mpp.mpg.de}
and S.~Vogl$^{\rm a,b\,}$\footnote{ stefan.vogl@physik.uni-freiburg.de}

\vspace{0.8cm}

{\em $^{\rm{a}}$Max-Planck-Institut 
f\"ur Physik (MPP)\\ F\"ohringer Ring 6,
80805 M\"unchen,
Germany} 
\\
\vspace{0.15 cm}
{\em $^{\rm{b}}$Albert-Ludwigs-Universität Freiburg, Physikalisches Institut\\
Hermann-Herder-Str. 3, 79104 Freiburg,
Germany}
\vspace*{0.8cm}

\end{centering}
\vspace*{0.3cm}
 
\noindent

\textbf{Abstract}:   The Kaluza-Klein (KK) decomposition of higher-dimensional gravity gives rise to a tower of KK-gravitons in the effective four-dimensional (4D) theory. Such massive spin-2 fields are known to be connected with unitarity issues and easily lead to a breakdown of the effective theory well below the naive scale of the interaction. However, the breakdown of the effective 4D theory is expected to be controlled by the parameters of the 5D theory. Working in a simplified Randall-Sundrum model we study the matrix elements for matter annihilations into massive gravitons. We find that truncating the KK-tower leads to an early breakdown of perturbative unitarity.
However, by considering the full tower we obtain a set of sum rules for the couplings between the different KK-fields that restore unitarity up to the scale of the 5D theory.  We prove analytically that these are fulfilled in the model under consideration and  present numerical tests of their convergence. This work complements earlier studies that focused on graviton self-interactions and yields additional sum rules that are required if matter fields are incorporated into warped extra-dimensions.

\vfill

\end{titlepage}
\setcounter{footnote}{0}

\begin{section}{Introduction}

Extra dimensions have been considered in physics for many different purposes since the seminal work of Kaluza \cite{Kaluza:1921tu} and Klein \cite{Klein:1926tv}. In particular from the early '90s onward   models featuring
large \cite{Antoniadis:1990ew,ArkaniHamed:1998rs,Appelquist:2000nn} and warped extra-dimensions \cite{Randall:1999ee,Randall:1999vf} have received considerable attention  due to their potential to resolve outstanding questions of the Standard Model such as the hierarchy problem.
However, even without a concrete particle physics model in mind, extra-dimensional models are interesting laboratories for the physics of massive spin-2 fields, see for example Sec.~10 of the review \cite{Hinterbichler:2011tt}. Most particle phenomenology inspired investigations of extra-dimensional models are chiefly interested in the role of the spin-2 fields as mediators between initial and final states consisting of matter fields \cite{Giudice:1998ck,Hooper:2007qk,Agashe:2006hk}. In contrast, studies motivated by a more theoretical interest in massive spin-2 fields have recently investigated the physics of KK-graviton scattering  \cite{Bonifacio:2019mgk,Chivukula:2019zkt,Chivukula:2019rij,Chivukula:2020hvi}. However, the production of spin-2 particles from matter has not received as much attention. This work aims to close this gap.
On the one hand, we expect our results to be relevant for phenomenological studies, for example for gravitationally interaction dark matter in extra-dimensional theories \cite{Lee:2013bua,Rueter:2017nbk,Folgado:2019sgz,Carmona:2020mmm}. On the other hand, such a study is also of interest from a more theoretical perspective since matter fields are a necessary ingredient in any realistic theory. In the following, we will focus on a concrete model of warped extra-dimension with two branes originally put forward by Randall and Sundrum
\cite{Randall:1999ee}.

Interactions between massive spin-2 fields and matter pose a subtle problem since scattering amplitudes involving them are plagued by unitarity issues, see \cite{Hinterbichler:2011tt} and references therein. The breakdown of the theory at high energies is already expected at the Lagrangian level but studies of the scattering amplitudes of spin-2 fields show a rapid growth in the high energy limit that indicates the break-down of perturbativity at scales much lower than the fundamental scale of the theory. This issue has received attention in the context of massive gravity \cite{ArkaniHamed:2002sp,Schwartz:2003vj,deRham:2010ik,deRham:2010kj} and the construction of theories that avoid this behavior is still being investigated \cite{Bonifacio:2019mgk,Gabadadze:2019lld}. In contrast, higher dimensional gravity is expected to be well-behaved up to the fundamental cut-off of the theory and, therefore, these issues should not arise in the associated 4D theories. Clearly, an individual KK-graviton cannot avoid the conclusions obtained from considerations of general spin-2 fields and, therefore, the other particles of the 4D theory have to be involved in the unitarization process that restores the fundamental scale of the underlying theory \cite{Schwartz:2003vj}. This is reminiscent of the unitarity problem in massive vector boson scattering in the Standard Model  which is resolved by including the Higgs boson \cite{Lee:1977eg}. The details of the cancellation mechanism depend on the geometry of the extra-dimension \cite{Schwartz:2003vj,Bonifacio:2019ioc} and are not know in general. However, the   
unitarization of KK-graviton scattering in warped extra-dimensions has recently been studied \cite{Chivukula:2020hvi}, see also \cite{Chivukula:2019zkt,Chivukula:2019rij}. So far, only  scattering of KK-gravitons has been considered.
However, a theory that describes a phenomenologically viable Universe also contains matter fields. We take a look at this 
previously neglected direction and investigate the origin and
the resolution of unitarity issues in processes connecting matter and KK-gravitons. For simplicity, we consider only a toy matter Lagrangian and include just a single fundamental scalar on the brane.
We analyze the matrix elements of scalar annihilations into KK-gravitons and find that unitarity is restored up to the fundamental scale once the full tower of KK-gravitons and the radion is included in the computation.
 Our study largely follows the approach of \cite{Chivukula:2020hvi} and we partially use their notation.

The paper is organized as follows. In Sec.~\ref{sec:RS_model} we briefly introduce warped extra-dimensions and comment on the connection between gravity in higher dimensions and the effective theory in 4D. Next, we analyze the matrix elements  scalar annihilations into final states consisting of KK-gravitons and radions. We pay close attention to the high energy behavior and identify sum rules involving the three-KK-graviton (and KK-graviton-radion) couplings required to restore perturbative unitarity up to the cut-off of the full theory. These sum rules are shown to be fulfilled in the RS-model both analytically and numerically in Sec.~\ref{sec:sum_rules}. Finally, we present our conclusions in~Sec.~\ref{sec:conclusions}

\end{section}

\begin{section}{The Randall-Sundrum model}
\label{sec:RS_model}

We analyze a simplified version of the Randall-Sundrum model \cite{Randall:1999ee} with a toy matter sector instead of the full Standard Model field content. To be concrete, our matter Lagrangian consists of a single scalar with only gravitational interactions. This setup is sufficient to make the point we are interested in and we expect that our key observations will carry over to a more realistic construction with minor modifications.  Our somewhat compressed introduction of the Randall-Sundrum model follows \cite{Chivukula:2020hvi}; for a more in-depth introduction see for example \cite{Rattazzi:2003ea,Kribs:2006mq,Raychaudhuri:2016kth}.

\begin{subsection}{The 5D Theory}

Before starting our discussion it is helpful to introduce some basic notation. 
We use capital Latin and lower case Greek letters, e.g. $M=0,1,2,3,4$  and $\mu=0,1,2,3$, to indicate 5-dimensional (5D) and 4-dimensional (4D)-indices, respectively. Thus, the coordinate of the full 5D space-time is denoted $x^M=(x^\mu,y)$.
The 5D space-time is compactified under an $S^1/\mathbb{Z}^2$ orbifold symmetry yielding a 5D bulk bounded by two 4-dimensional (4D) branes located at $y=0$ and $y=\pi r_c$ where $y$ indicates the coordinate of the fifth dimension and $r_c$ its size.
 This compactification symmetry  leads to the identification $(x,y)=(x,-y)$ which allows to extend the coordinate range to $y\in [-\pi r_c, \pi r_c]$.
 It is often convenient to work with dimensionless quantities instead of dimensional ones which can be achieved by normalizing with respect to $r_c$,  e.g. $\varphi=y/r_c$. Gravity permeates the bulk while matter fields are taken to be localized on the branes.

The action of the theory is given by 
\begin{equation}
        S = S_{\text{bulk}}+S_{\text{UV}}+S_{\text{IR}} \ ,
    \end{equation}
with
    \begin{align}
    \label{eq:RS_Action}
            & S_{\text{bulk}} = \frac{1}{2}M_5^3\int d^4x \int\limits_{-\pi}^\pi d\varphi \sqrt{G}(R-2\Lambda_B ) \nonumber  ,\\
            & S_{\text{UV}} =\int d^4x\int\limits_{-\pi}^\pi d\varphi \sqrt{-g_{\text{UV}}}(-V_{\text{UV}}+\mathcal{L}_{\text{UV}})\delta(\varphi) \nonumber ,\\
            & S_{\text{IR}}=\int d^4x\int\limits_{-\pi}^\pi d\varphi \sqrt{-g_{\text{IR}}}(-V_{\text{IR}}+\mathcal{L}_{\text{IR}})\delta(\varphi-\pi) \ ,
    \end{align}
where $G$ is the determinant of the 5D metric, $R$ the Ricci scalar and $M_5$ the 5D Planck mass.  $\Lambda_B$ denotes the vacuum energy of the bulk while $V_{\text{UV}}$ and $V_{\text{IR}}$ are the vacuum energy terms on the brane.  $\mathcal{L}_{\text{IR}}$ and $\mathcal{L}_{\text{UV}}$ are the Lagrange densities of fields that are localized to the 4D branes while $g_{\text{IR/UV}}$ are the 4D metric on the respective branes. For simplicity we will take $\mathcal{L}_{\text{UV}}=0$ and 
\begin{align}
    \mathcal{L}_{\text{IR}}= \frac{1}{2} \partial_\mu \phi \partial^\mu \phi -\frac{1}{2} m^2 \phi^2
\end{align}
where $\phi$ is a scalar field without any interactions besides gravity. Neglecting the matter part, Einstein's equation is solved by the metric
\begin{equation}
\label{eq:metric}
    G_{MN}= 
    \begin{pmatrix}
     w(x,y) g_{\mu\nu} & 0\\
     0 & -v(x,y)^2
    \end{pmatrix}\,.
\end{equation}
Choosing the vacuum energy contributions on the branes and in the bulk such that  the solution respects 4D Poincaré invariance allows to fix $w(x,y)$ and $v(x,y)$ and leads to an invariant distance interval 
\begin{equation}
    \text{ds}^2= e^{-2k |y|}\eta_{\mu\nu} \text{d}x^\mu \text{d}x^\nu- \text{d}y^2 \ ,
\end{equation}
where $\eta_{\mu \nu}= \mbox{Diag}(+1,-1,-1,-1,)$ is the flat metric in 4D and $k$ denotes the  warping  parameter defined as $k \equiv \sqrt{\frac{-\Lambda_B}{6}}$. It should be noted that in order to ensure 4D Poincaré invariance, the branes' vacuum energies are constrained to be $    V_{\text{UV}}=-V_{\text{IR}}=6M_5^3k$.  
By performing the integral over the 5th dimension in eq.~\ref{eq:RS_Action} we can re-express the theory in terms of an effective Lagrangian in 4D.
In the usual Randall-Sundrum model this allows to alleviate the hierarchy problem of the SM since the vacuum expectation value (vev) of the Higgs $v$ defined in 5D is related to the one in 4D by the warping factor $e^{-k \pi r_c}$. 
For convenience, we define the dimensionless parameter
$\mu=k r_c$ to simplify the exponent of the warp factor. 
For values of $ \mu \approx 12$ 
the exponential factor allows for a TeV scale vev even if all fundamental mass-dimensional parameters of the 5D theory are $\mathcal{O}(M_{Pl})$, thus resolving the hierarchy problem. As we will not consider the SM explicitly we do not have a preference for a specific value of the warp factor but we will focus on the limit~$e^{-\mu\pi} \ll 1$.  

The gravitational field content of the Randall-Sundrum model is obtained through a weak-field expansion of the metric around the vacuum solution, i.e. 
\begin{equation}
     G_{MN} \quad  \longmapsto	\quad
 G_{MN}+\kappa \ h_{MN} \ ,
\end{equation}
where $\kappa$ is an expansion parameter defined as $\kappa =2/M_5^{3/2}$.
The expansion generates scalar, vector and tensor perturbations, corresponding to $h_{\mu\nu}$, $h_{\mu4}$ and $h_{44}$, respectively. 
In the Randal-Sundrum model it is possible to choose the gauge such that the vector component vanishes even though this does not hold for general higher dimensional models \cite{Callin:2004zm}. 
The tensor perturbation correspond to a spin-2 field, i.e. a 5D-graviton, while the scalar perturbation, the radion, is related to the width of the 5th dimension. 
We utilize the Einstein frame parameterization~\cite{Csaki:2000zn} which amounts to the following replacement in Eq.~\ref{eq:metric}
\begin{equation}
    w(x,y)= e^{-2(k|y|+\hat{u})}\quad , \quad v(x,y)= 1+2\hat{u} \quad  \ ,
\end{equation}
where $\hat{u}$ contains the radion field. This ansatz eliminates the mixing between the radion and the gravitons. We take $g_{\mu\nu}$ to be weakly perturbed around a flat background
\begin{align} 
g_{\mu\nu} = \eta_{\mu\nu}+\kappa \hat{h}_{\mu\nu} \ ,
\end{align}
where $\hat{h}_{\mu\nu}$ denotes a symmetric tensor field
that includes the graviton. The metric is then given by
\begin{equation}
    G_{MN}= 
    \begin{pmatrix}
     e^{-2(k|y|+\hat{u})}(\eta_{\mu\nu}+\kappa \ \hat{h}_{\mu\nu}) & 0\\
     0 & -(1+2\hat{u})^2
    \end{pmatrix} \ .
\end{equation}
Denoting the radion field $\hat{r}$, we define $\hat{u}$ as
\begin{equation}
    \hat{u}(x,y) \equiv \kappa \ \frac{\hat{r}(x)}{2\sqrt{6}} \ e^{2k|y|} \ ,
\end{equation}
where the fact that the $y$ dependence of $\hat r$ can be removed by an appropriate choice of coordinates~\cite{Callin:2004zm} has been employed. 
By expanding the full Lagrangian of the theory in powers of $\kappa$ we obtain, order by order, a theory of interacting 5D graviton and radion fields $\hat h_{\mu\nu}$ and $\hat r$. We expand the bulk Lagrangian to third power in the fields since this includes the three-graviton interaction Lagrangian that is crucial for our studies.  In addition, we need the first two interactions between the scalar field $\phi$ and the gravitons and radions. The key results of this expansion are summarized in Appx.~\ref{Appx:Interactions}.
\end{subsection}

\begin{subsection}{Effective theory in 4D}
By integrating out the 5th dimension, this model can be reduced to an effective theory in 4D. To achieve this, we employ the Kaluza-Klein (KK) decomposition of the 5D fields
\begin{equation}
   \begin{split}
        \hat{h}_{\mu\nu}(x,y)&=\sum\limits_{n=0}^\infty\frac{1}{\sqrt{r_c}}h^{(n)}_{\mu\nu}(x)\ \psi_n(\varphi(y)) \ ,\\
    \hat{r}(x)&=\frac{1}{\sqrt{r_c}} \psi_r \ r(x) \ ,
   \end{split}
\end{equation}
where $\psi_{n}(\varphi)$ absorb the 5D-dependence of the fields and the unhatted $h$ and $r$ fields carry the $x$ dependence. As indicated above, $\psi_r$ is independent of $\varphi$. 
The decomposition transforms the single 5D-graviton into a tower of 4D-gravitons.
In order to get the canonical massive Fierz-Pauli Lagrangian for the gravitons~\cite{Fierz:1939ix}, the 5D-components of the KK-decomposition must satisfy the following differential equation~\cite{Davoudiasl:1999jd}
\begin{equation}
\label{eq:mass-phi}
    \frac{1}{r_c^2} \der{}{\varphi}\left[A(\varphi)^4\der{\psi_n}{\varphi}\right]=-m_n^2 A^2 \psi_n \ ,
\end{equation}
where we have introduced the shorthand $A(\varphi)=e^{-\mu|\varphi|}$ and $m_n$ is the mass of the $n$-th graviton.
This equation is a particular case of the more general Sturm-Liouville equation. It can be proved that $m_n \in \mathbb{R}$ with $m_{n}<m_{n+1}$ and
the solutions $\psi_n(\varphi)$ are orthogonal and normalized with respect to the scalar product
\begin{equation}
\label{eq:orthogonality}
      \left<\psi_n,\psi_m \right> =\int\limits_{-\pi}^{\pi} d\varphi \ A(\varphi)^2 \ \psi_n(\varphi)\psi_m(\varphi)=\delta_{n,m} \ .
\end{equation}
Consistency with a phenomenological acceptable 4D gravity requires the graviton with $n=0$ to correspond to the massless graviton of General Relativity and, hence, $m_0=0$ and $\psi_0 = \text{constant}$. $\psi_0$ and $\psi_r$ are fixed by the normalization to
\begin{equation}
    \psi_0 = \sqrt{\frac{\mu}{1-e^{-2\mu\pi}}} \simeq  \sqrt{\mu} \quad , \quad \psi_r = \sqrt{\frac{\mu }{e^{2 \pi  \mu }-1}} \simeq  \sqrt{\mu} \ e^{-\mu\pi}\,.
\end{equation}
The functions $\psi_{n>0}(\varphi)$ and the masses can be determined by solving Equation \ref{eq:mass-phi} with the boundary conditions $\partial_\varphi\psi_n|_{\varphi=0,\pi}=0$. 
In the limit $e^{-\mu \pi} \ll 1$ the solutions simplify and can be approximated as
\begin{equation}
     \psi_n(\varphi)\simeq \frac{e^{2\mu |\varphi|}}{N_n}J_2\left(\gamma_n e^{\mu(|\varphi|-\pi)}\right) \quad \mbox{and}\quad   m_n \simeq k \gamma_n e^{-\mu\pi} \ ,
\end{equation}
where $J_i$ denoted the $i$th Bessel $J$-function while  $\gamma_n$ is the $n$th zero of $J_1(x)$. The normalization factors $N_n$ are given by
\begin{equation}
    N_n \simeq - \frac{e^{\mu\pi}}{\sqrt{\mu}}J_0(\gamma_n) \ .
\end{equation}

Performing the integration over the 5th dimension on the quadratic pieces of the pure gravity Lagrangian yields the kinetic terms of a massless spin-2 field, i.e. the graviton of general relativity, a massless spin-0 field, i.e. the radion, and an infinite number of spin-2 fields with Fierz-Pauli mass terms, i.e. a tower of massive KK-gravitons. Decomposing the first order weak field expansion of the matter Lagrangian leads to following interaction between matter and gravitons
\begin{equation}
    \mathcal{L}_{\text{int}}^{(1)}=-\frac{1}{2}\kappa T_{\mu\nu}\hat{h}^{\mu\nu}(x,\varphi=\pi)=-\frac{1}{2}\kappa T_{\mu\nu}\left(\sum\limits_{n=0}^\infty \frac{1}{\sqrt{r_c}}h_n^{\mu\nu}(x)\psi_n(\pi)\right) \ .
\end{equation}
Requiring that the massless graviton matches the expectation from GR allows to fix the relation between the (reduced) Planck mass in 4D, $M_{Pl}$, and the parameters of the 5D theory 
\begin{equation}
    \frac{1}{2}\kappa \frac{1}{\sqrt{r_c}}\psi_0 =\frac{1}{M_{Pl}} \quad \mbox{or, equivalently,} \quad M_{Pl}^2 = \frac{M_5^3}{k}\left(1-e^{-2\mu\pi}\right) \ ,
\end{equation}
 which simplifies to $  M_5^3 \simeq k M_{Pl}^2$ in the limit $e^{-\mu\pi}\ll1$. Due to the different normalization, the strength of the interaction of the other KK-fields $h^{\mu\nu}_{n>0}$  
 is controlled by a combined scale $\Lambda$ defined by $\Lambda^{-1}= M_5^{-3/2} \psi_n(\pi)/\sqrt{r_c}$ which leads to $\Lambda \simeq M_{Pl}\, e^{-\mu \pi}$ in the large $\mu$ limit. The radion contribution to the interaction Lagrangian is
\begin{equation}
    \mathcal{L}_{\text{int,r}}^{(1)}=\frac{1}{\sqrt{6}\Lambda}rT \ ,
\end{equation}
where $T=\eta^{\mu\nu}T_{\mu\nu}$ is the trace of the energy-momentum tensor of the matter field. 
 In the 4D reduction of the higher powers of the expanded Lagrangian, interactions between all combinations of massless graviton, massive KK-modes and the radion with matter appear. The strength of these interactions is given by a generalized scale $\Lambda_{n0,n_m,n_r}^N= \Lambda^{n_m+n_r} M^{n_0}_{Pl} $ 
where  $n_{0}$ ($n_{m}$) is the number of massless (massive) gravitons, $n_r$ the number of radions and $N=n_0+n_m+n_R$. 

In addition to the interactions between matter and the gravitons or radions we also need the cubic interactions between these fields which are substantially more involved. Instead of considering a handful of simple interactions on the brane we now need to treat the interactions between all constituents of the KK-tower. The strength of the interaction is controlled by the  overlap of the wave functions or their derivatives in the bulk. As it will turn out to be helpful later, we first introduce a more general parameterization of the coupling coefficients than strictly needed to define the interactions in 4D\footnote{The $a$-$b$ notation is inspired by~\cite{Chivukula:2020hvi}, however, we prefer to normalize our coefficients differently.}. 
The 5D Lagrangian consists of contributions that possess either no or two 5D derivatives.  
The interaction between 3-gravitons  does not contain any derivatives and we can define a coefficient $a$ parameterizing the wave-function overlap. In addition, we can also define a different class of coefficients $b$ that include two 5D derivatives\footnote{The $b$ type integrals do not appear directly in the Lagrangian but they will turn out to the useful later.}. Labeling the fields from $n_1$ to $n_3$ the $a$'s and $b$'s read
\begin{equation}
\label{eq:coefficients}
    \begin{split}
         a_{\Vec{n}}&:=\int\limits_{-\pi}^{\pi} \diff{\varphi} \ A(\varphi)^2 \ \psi_{n_1}(\varphi)\psi_{n_2}(\varphi)\psi_{n_3}(\varphi)\ ,\\
         b_{\Vec{n}}&:=\int\limits_{-\pi}^{\pi} \diff{\varphi} \ A(\varphi)^4 \ \psi'_{n_1}(\varphi)\psi'_{n_2}(z)\psi_{n_3}(\varphi) \ ,
    \end{split}
\end{equation}
where $\vec{n}=(n_1,n_2,n_3)$ indicates the set of fields involved. As can be seen from the definition, $a$-type coefficients are symmetric under permutation of all indices while $b$-type coefficients are only symmetric under permutation of the first two fields.
\begin{center}
\begin{minipage}[h]{.1\textwidth}
 \begin{figure}[H]
      \includegraphics[scale=0.35]{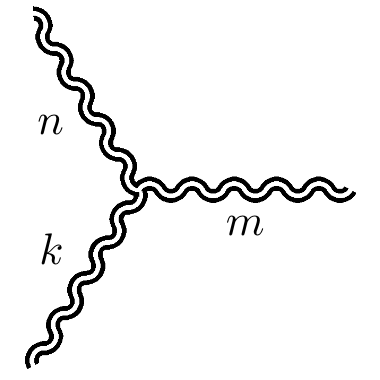}
\end{figure}
\end{minipage}
\begin{minipage}[h]{.1\textwidth}
\begin{flalign*}
   \quad\quad\propto \frac{a_{\text{nkm}}}{\sqrt{r_c}M_5^{3/2}} &&
    \end{flalign*}
\end{minipage}
\begin{minipage}[h]{.1\textwidth}
 \begin{figure}[H]
       \includegraphics[scale=0.35]{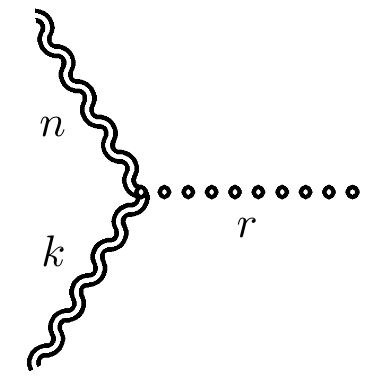}
\end{figure}
\end{minipage}
\begin{minipage}[h]{.1\textwidth}
\begin{flalign*}
   \quad\quad\propto \frac{   b_{\text{nkr}} }{r_c^{5/2}M_5^{3/2} } &&
    \end{flalign*}
\end{minipage}
\begin{minipage}[h]{.1\textwidth}
 \begin{figure}[H]
       \includegraphics[scale=0.35]{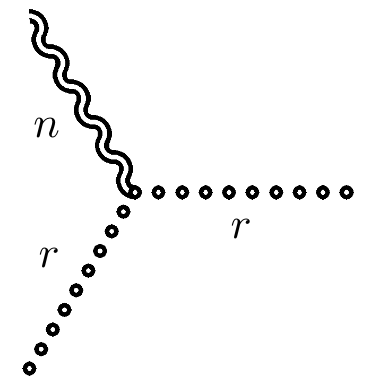}
\end{figure}
\end{minipage}
\begin{minipage}[h]{.24\textwidth}
\begin{flalign*}
   \quad\quad \propto \frac{  c_{\text{nrr}}}{ \sqrt{r_c}M_5^{3/2} }  &&
    \end{flalign*}
\end{minipage}
    \captionof{figure}{Cubic interactions between gravitons and radions.}
    \label{vertices--}
\end{center}
In addition, there are processes involving radion fields. In the two-graviton-radion vertex, a second kind of $b$-coefficients appear since the relevant Lagrangian includes two 5D derivatives. One finds 
\begin{align}
    b_{n_1,n_2,r}:= \int\limits_{-\pi}^{\pi} \diff{\varphi} \ A(\varphi)^2 \ \psi'_{n_1}(\varphi)\psi'_{n_2}(z)\psi_{r} 
\end{align}
which is also symmetric in the first two indices. The two-radion graviton vertex is simpler. We denote the coefficient $c_{nrr}$ and find that it is given by
\begin{align}
   c_{nrr}:= \int\limits_{-\pi}^{\pi} \diff{\varphi} \ A(\varphi)^{-2} \ \psi_{n}(\varphi)\psi_{r}\psi_{r} 
\end{align}
Integrating out the 5D and considering all the powers of $M_5$ and of $r_c$ deriving from the Lagrangian expansion and the KK-decomposition, the strength of the cubic interactions in the 4D theory are given by the vertices of Fig. \ref{vertices--}.
In the large $\mu$-limit these expressions simplify and we can split off the $\mu$ dependence,  which combines with $M_5$ to set the overall scale of the interactions. This simplifies the integrals over the fifth dimension, which separate into numerical constants that do not depend on the parameters of the theory any more, and into powers of $k e^{-\mu\pi} \simeq m_n/\gamma_n$. In this case we get
\begin{align}
\label{eq:couplings_1}
 \frac{a_{knm}}{\sqrt{r_c}M_5^{3/2}}&\longrightarrow \frac{\chi_{knr} \sqrt{\mu} e^{ \mu\pi}}{\sqrt{r_c}M_5^{3/2}}=\frac{\chi_{knm}}{\Lambda}  \\
 \label{eq:couplings_2}
  \frac{b_{knr}}{r_c^{5/2}M_5^{3/2}}&\longrightarrow \frac{\tilde{\chi}_{knr} \mu^{5/2} e^{ -\mu\pi}}{r_c^{5/2} M_5^{3/2}}=\frac{\tilde{\chi}_{knr}}{\Lambda}\left(k^2e^{-2\mu\pi}\right) \\
  \label{eq:couplings_3}
  \frac{c_{nrr}}{\sqrt{r_c}M_5^{3/2}}&\longrightarrow \frac{\chi_{nrr} \sqrt{\mu} e^{ \mu\pi}}{\sqrt{r_c}M_5^{3/2}}=\frac{\chi_{nrr}}{\Lambda} 
\end{align}
where the numerical coefficients are given by
\begin{equation}
    \begin{split}
        \chi_{nkj}&\equiv \frac{-2}{J_0(\gamma_n)J_0(\gamma_k)J_0(\gamma_j)}\int\limits_0^1 \diff{u} \ u^3 J_2(\gamma_nu)J_2(\gamma_ku)J_2(\gamma_ju)\ ,\\
         \tilde{\chi}_{knm}&\equiv 2\frac{\gamma_n \gamma_k}{J_0(\gamma_n) J_0(\gamma_k)}\int\limits_0^1 \diff{u} \ u^{3}  J_1\left(\gamma_n u\right) J_1\left(\gamma_k u\right)\ ,\\
          \chi_{nrr}&\equiv \frac{-2}{J_0(\gamma_n)}\int\limits_0^1 \diff{u} \ u^3 J_2(\gamma_nu)=-\frac{2 J_3\left(\gamma_{n}\right)}{\gamma_{n} J_0\left(\gamma_{n}\right)}\ .
    \end{split}
\end{equation}
Comparing Eq. \ref{eq:couplings_1}, \ref{eq:couplings_2} and \ref{eq:couplings_3}, it can be seen that the term with the $b_{nkr}$ coupling does not have the same energy-dimension as the other two. This follows directly from the presence of 5D-derivatives which give rise to this term and that are absent in the other two. Consequently, in the 4D Lagrangian, the $a_{knm}$ and $c_{nrr}$ couplings are multiplied either by masses of the gravitons or by their momenta while the term proportional to  $b_{knr}$ just contains combinations of the flat metric $\eta_{\mu\nu}$.

For completeness we also list the coefficient of the interaction between three radions $\chi_{rrr}$.  In this case the integration over the fifth dimension is trivial and we find
\begin{align}
              \chi_{rrr}&\equiv 2\int\limits_0^1 \diff{u} \ u^3= \frac{1}{2}\ .
\end{align}
The numerical couplings above can be generalized to a higher number of participating interacting particles by inserting in the integrals a factor of $u^2$ for every radion, and a factor of $-u^2\frac{J_2(\gamma_j u)}{J_0(\gamma_j)}$ for every $j$-graviton. 
We report the Feynman rules for all relevant interactions in the large $\mu$ limit in Appx.~\ref{Appx:Feynman}. 
\end{subsection}

\end{section}

\begin{section}{Matrix elements for graviton production}
In order to assess the validity of the theory we analyze the matrix elements of processes that involve the KK-gravitons. We focus on two classes of interactions: 1) KK-graviton production in $\phi$ annihilations ($\phi \phi \rightarrow G_n G_m$) and 2) mixed graviton-radion production ($\phi \phi \rightarrow G_n r$).
The first class is most interesting from a theoretical point of view and will allow us to identify new sum rules for the couplings of three gravitational fields while the second class leads to an independent relation for the radion couplings. 

\subsection{Helicity matrix elements for $\phi \phi \rightarrow G_k G_n$}

\begin{figure}[tb]
\centering
\includegraphics[width=0.7\textwidth]{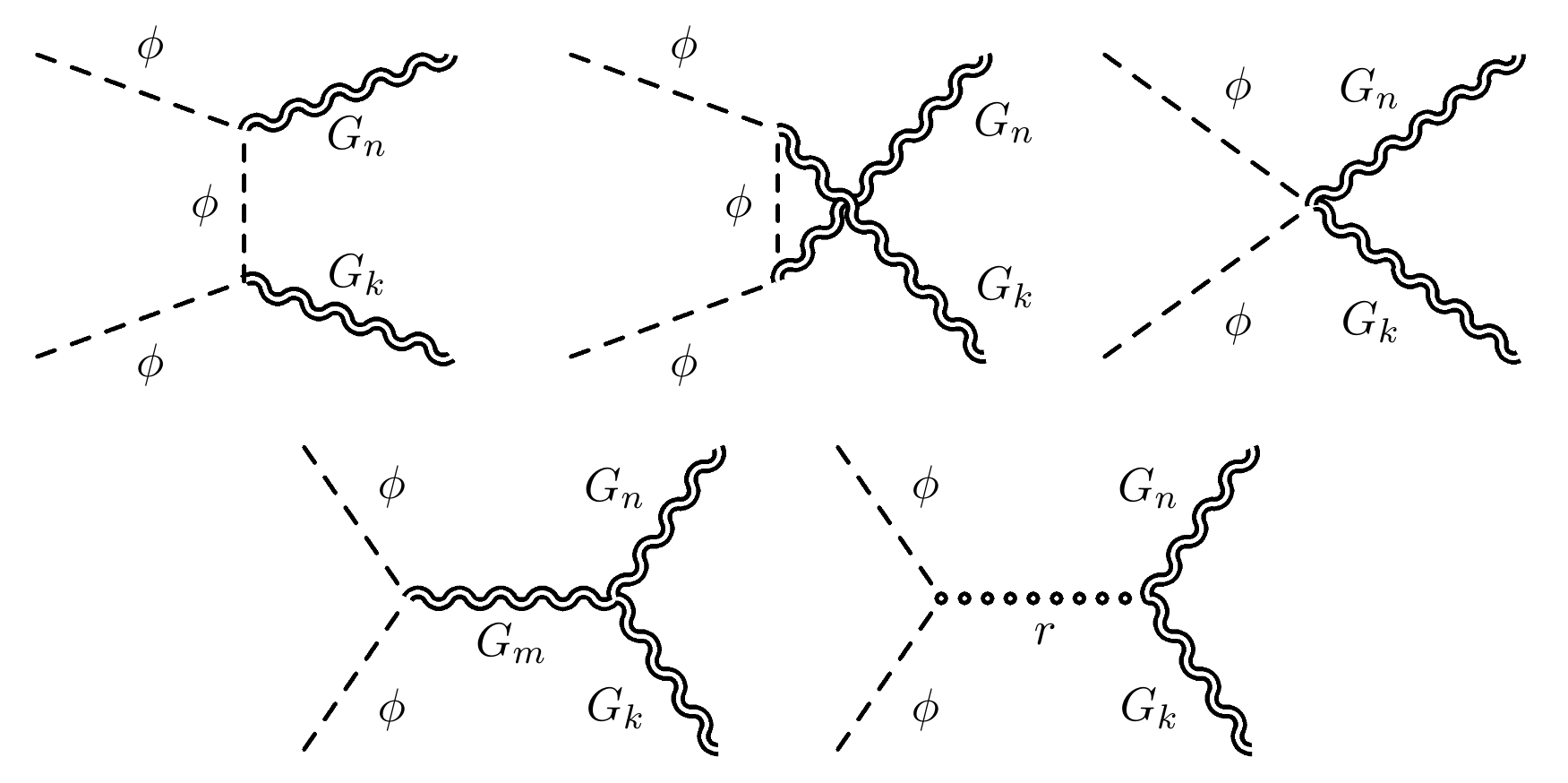}\par
\caption{Representative set of diagrams contributing to $\phi\phi \rightarrow G_k G_n$ annihilations.}\label{fig:GravProd}
\end{figure}
We are interested in the high energy limit of the $\phi\phi \rightarrow G_n G_k$ matrix element.
There are four classes of diagrams that contribute. First, there is the emission of $G$ via t- and u-channel $\phi$ exchange and a contribution from the $\phi \phi  G_n G_k$ contact interaction. In addition, there is an infinite set of diagrams with all possible graviton-modes in the s-channel and one diagram from s-channel radion exchange. A representative set of diagrams is shown in Fig.~\ref{fig:GravProd}.

We work in the center of mass frame and take the initial state particle to travel along the $z$-axis while the final state particle are emitted back to back in some arbitrary direction.
Using the helicity formalism for spin-2 fields,  the spin-2 polarization tensor can be decomposed as a combination of spin-1 polarization tensors \cite{Gleisberg:2003ue}:
\begin{equation}
   \begin{split}
       \epsilon_{0}^{\mu\nu} & = \frac{1}{\sqrt{6}}\left( \epsilon_{\pm 1}^{\mu}\epsilon_{\mp 1}^{\nu}+2 \epsilon_{0}^{\mu}\epsilon_{0}^{\nu}+ \epsilon_{\mp1}^{\mu}\epsilon_{\pm 1}^{\nu}\right) \ ,\\
       \epsilon_{\pm 1}^{\mu\nu} & = \frac{1}{\sqrt{2}}\left(\epsilon_{\pm 1}^{\mu}\epsilon_{0}^{\nu}+ \epsilon_{0}^{\mu}\epsilon_{\pm 1}^{\nu}\right)\ , \\
       \epsilon_{\pm 2}^{\mu\nu} & = \epsilon_{\pm 1}^{\mu}\epsilon_{\pm 1}^{\nu} \ .\\
   \end{split}
\end{equation}
 An explicit form for the spin-1 polarization tensor of a massive particle moving in an arbitrary direction, i.e. a particle with with mass $m$ and momentum $\Vec{p}=|\vec{p}|\left(\sin\theta\cos\phi,\sin\theta\sin\phi,\cos\theta \right)$,  is given by
\begin{equation}
\label{pol-tens-exp}
\begin{split}
       \epsilon_{\pm 1}^{\mu} & = \frac{1}{\sqrt{2}}e^{\mp i \gamma}\left(0;\mp \cos\theta\cos\phi + i \sin\phi,\cos\theta\sin\phi -i \cos\phi,\pm \sin\theta \right) \ , \\
       \epsilon_{0}^{\mu} & =\frac{E}{m}\left(\sqrt{1-\frac{m^2}{E^2}};\sin\theta\cos\phi,\sin\theta\sin\phi,\cos\theta\right)\ .\\
\end{split}
\end{equation}
Since we only consider scalar particles in the initial state $\phi$ and $\gamma$ can be chosen to be zero.

In the following, we neglect $m_\phi$ and focus on the high energy limit of the annihilations. We expand the amplitudes in powers of the center of mass energy $\sqrt{s}$. Since we are dealing with an effective theory and the interaction vertex comes with a suppression scale $\Lambda$, it is clear that we have to find contributions to the matrix elements with $\mathcal{M}\propto s$. However, looking at the polarization vectors of the longitudinal modes we observe an additional growth for $E\gg m$. Thus we expect to find contributions that grow even faster with $s$. Truncating the s-channel diagrams after the first graviton this is indeed the case and we find contributions with $\mathcal{M} \propto s^3$. 
 This is a bit of a  puzzle since this would imply a breakdown of the theory well below the fundamental scale. 
However, there is no guarantee that a truncated 4D theory will respect the properties of the 5D theory. Therefore, we expect that the anomalous growth with higher powers of $s$ cancels once the full theory, i.e. the untruncated KK-tower, is included in the analysis \footnote{Recently, this was shown explicitly for the elastic scattering of gravitons, which exhibits an even worse high energy behavior  \cite{Chivukula:2020hvi}. This is resolved by sum rules relating the 3-point interactions of the KK-gravitons to the 4-point interactions. In contrast, our amplitudes do not depend on the 4-point interactions and our relations only involve the three-KK-graviton coupling.}.

We now present the expanded matrix elements in the large $s$ limit order by order in $\sqrt{s}$. This allows us to identify sum rules for the couplings of the theory that ensure the cancellation of contributions that grow faster than $s$.
We will demonstrate explicitly that these sum rules are fulfilled in the Randall-Sundrum model in section \ref{sec:sum_rules}. 
 Note that we show only the final state helicities that are non zero at a given order. In general, we find that the amplitudes related to 0-helicity states are the most relevant since they include two of the longitudinal graviton modes that get enhanced in the high energy limit.
 
{\bf Order $\bf s^3$:} At this order only the helicity zero final state contributes. We find 
\begin{align}
\mathcal{M}(0,0)=-\frac{i s^3 \left(  \sin ^2(\theta)\right) \left(\sum_m \chi _{\text{nkm}}-1\right)}{24 \Lambda ^2 m_k^2 m_n^2} + \mathcal{O}(s^2)
\end{align}
As can be seen the $\mathcal{O}(s^3)$ contribution vanishes only if 
\begin{align}
\sum\limits_{m=1}^\infty \chi_{nkm}=1\;.
\label{eq:sum_s3}
\end{align} This is indeed the case in the Randall-Sundrum model and we will prove it analytically in Section \ref{sec:sum_rules}.

{\bf Order $\bf s^2$:} Here we find more final state helicities that contribute to the expanded matrix element, see Tab.~\ref{Tab:Amplitudes_s2}.
\begin{table}[t]
\begin{center}
\begin{tabular}{cc|c}
\toprule
$\lambda_n$&$\lambda_k$& Amplitude \\ 
\midrule
-2 & 0 & $-\frac{i s^2 \sin ^2(\theta ) \left(\sum_m\chi _{\text{nkm}}-1\right)}{4 \sqrt{6} \Lambda ^2 m_k^2}$\\
-1 & -1 & $-\frac{i s^2 \sin ^2(\theta ) \left(\sum_m\chi _{\text{nkm}}-1\right)}{4 \Lambda ^2 m_k m_n}$ \\
0 & -2 & $-\frac{i s^2 \sin ^2(\theta ) \left(\sum_m\chi _{\text{nkm}}-1\right)}{4 \sqrt{6} \Lambda ^2 m_n^2}$ \\
0 & 0 & \text{see eq.~\ref{eq:sum_s2}}\\
0 & +2 & $-\frac{i s^2 \sin ^2(\theta ) \left(\sum_m\chi _{\text{nkm}}-1\right)}{4 \sqrt{6} \Lambda ^2 m_n^2}$ \\
+1 & +1 & $-\frac{i s^2 \sin ^2(\theta ) \left(\sum_m\chi _{\text{nkm}}-1\right)}{4 \Lambda ^2 m_k m_n}$  \\
+2 & 0 &$-\frac{i s^2 \sin ^2(\theta ) \left(\sum_m\chi _{\text{nkm}}-1\right)}{4 \sqrt{6} \Lambda ^2 m_k^2}$ \\
\bottomrule
\end{tabular}
\end{center}
\caption{Contribution to the matrix element for the production of the $k$-th and $n$-th gravitons with relative helicities $\lambda_n$ and $\lambda_k$ at $\mathcal{O}(s^2)$. \label{Tab:Amplitudes_s2}}
\end{table}
As can be seen all final states except the  
$(0,0)$-helicity state vanish if the sum rule inferred from the $s^3$ contribution to $\mathcal{M}$ is fulfilled. The $(0,0)$-helicity state is more complicated and we find
\begin{equation}
\begin{split}
\mathcal{M}(0,0)=-\frac{i s^2}{144 \Lambda ^2 m_k^2 m_n^2} & \left[(m_k^2+m_n^2) \left(3 \cos (2 \theta )( 4\sum_m\chi_{\text{nkm}}-5)+(4\sum_m\chi_{\text{nkm}}-9)\right)  \right.\\
& \left. +4 \sum_m\chi _{\text{nkm}} \left(m_k^2-m_n^2\right)^2/m_m^2+(3 \cos (2 \theta ) +1) \sum_m m_m^2\chi _{\text{nkm}} \right.\\
& \Bigl. +24 \cos ^2(\theta ) m_{\phi }^2 \left(\sum_m\chi_{\text{nkm}}-1\right)+24 k^2e^{-2\mu\pi} \tilde{\chi }_{\text{nkr}}\Bigr]\\
&  + \mathcal{O}(s^{3/2})\ .
\end{split}
\label{eq:sum_s2}
\end{equation} 
In contrast to the previous case, this expression also depends on $\tilde\chi_{nkr}$, i.e. the radion contribution is crucial for the cancellation. The different angular dependence allows to identify two separate sum-rules at this order.
The part proportional to $\cos(2\theta)$ vanishes only if 
\begin{align}
\sum_{m=1}^\infty \chi_{nkm} \ m_m^2 =m_n^2+m_k^2
\label{eq:sum_s2_1}
\end{align}
while the remaining part requires
\begin{align}
\sum\limits_{m=1}^\infty\frac{\chi_{nkm}}{m_m^2}(m_n^2-m_k^2)^2-(m_n^2+m_k^2)+6\ k^2e^{-2\mu\pi}\tilde{\chi}_{nkr} = 0\,.
\end{align}
While these sum rules look more daunting than eq.~\ref{eq:sum_s3}, it can be shown that they hold in the model under consideration.

{\bf Order $s^{3/2}$:} Finally, we find contributions to the matrix element that scale as $s^{3/2}$ that are summarized in tab.~\ref{tab:amplitudes_s32}.
\begin{table}[t]
\begin{center}
\begin{tabular}{cc|c}
\toprule
$\lambda_n$&$\lambda_k$& Amplitude \\ 
\midrule
-2 & -1 & $\frac{i s^{3/2} \sin (2 \theta ) \left(\sum_m\chi _{\text{nkm}}-1\right)}{4 \Lambda ^2 m_k}$ \\
-1 & -2 & $\frac{i s^{3/2} \sin (2 \theta ) \left(\sum_m\chi _{\text{nkm}}-1\right)}{4 \Lambda ^2 m_n}$\\
-1& 0 & $-\frac{i s^{3/2} \sin (2 \theta ) \left((m_k^2+m_n^2 ) \left(\sum_m\chi _{\text{nkm}}-2\right)+\sum_m m_m^2 \chi _{\text{nkm}}\right)}{8 \sqrt{6} \Lambda ^2 m_k^2 m_n}$ \\
0 & -1 & $-\frac{i s^{3/2} \sin (2 \theta ) \left((m_k^2+m_n^2 ) \left(\sum_m\chi _{\text{nkm}}-2\right)+\sum_m m_m^2 \chi _{\text{nkm}}\right)}{8 \sqrt{6} \Lambda ^2 m_k m_n^2}$ \\
0 & +1 & $\frac{i s^{3/2} \sin (2 \theta ) \left((m_k^2+m_n^2 ) \left(\sum_m\chi _{\text{nkm}}-2\right)+\sum_m m_m^2 \chi _{\text{nkm}}\right)}{8 \sqrt{6} \Lambda ^2 m_k m_n^2}$ \\
+1 & 0 & $\frac{i s^{3/2} \sin (2 \theta ) \left((m_k^2+m_n^2 ) \left(\sum_m\chi _{\text{nkm}}-2\right)+\sum_m m_m^2 \chi _{\text{nkm}}\right)}{8 \sqrt{6} \Lambda ^2 m_k^2 m_n}$\\
+1 & +2 & $-\frac{i s^{3/2} \sin (2 \theta ) \left(\sum_m \chi _{\text{nkm}}-1\right)}{4 \Lambda ^2 m_n}$\\
+2 & +1 & $-\frac{i s^{3/2} \sin (2 \theta ) \left(\sum_m \chi _{\text{nkm}}-1\right)}{4 \Lambda ^2 m_k}$ \\
\bottomrule
\end{tabular}
\end{center}
\caption{Contribution to the amplitudes for the production of the $n$-th and $k$-th gravitons with relative helicities $\lambda_n$ and $\lambda_k$ at $\mathcal{O}(s^{3/2})$.\label{tab:amplitudes_s32}}
\end{table}
Clearly, four of the eight non-zero entries in the table vanish if the $\mathcal{O}(s^3)$-sum rule holds.
The remaining contributions (containing a 0-mode), are all proportional to 
\begin{equation}
   \propto \left((m_k^2+m_n^2 ) \left(\sum_m\chi _{\text{nkm}}-2\right)+\sum_m m_m^2 \chi _{\text{nkm}}\right) \ .
\end{equation}
After imposing eq.~\ref{eq:sum_s3} this just reduces to eq.~\ref{eq:sum_s2_1}. Therefore, all contributions at this order vanish if the sum rules derived for the contributions at higher power in $s$ hold. 

\subsection{Helicity matrix elements for $\phi \phi \rightarrow G_n r$}

Now we turn towards the the annihilation of DM-particles into a graviton and a radion. A representative set of diagrams is shown in Figure \ref{radGravProd}. 
\begin{figure}[tb]
\centering
\includegraphics[width=0.7\textwidth]{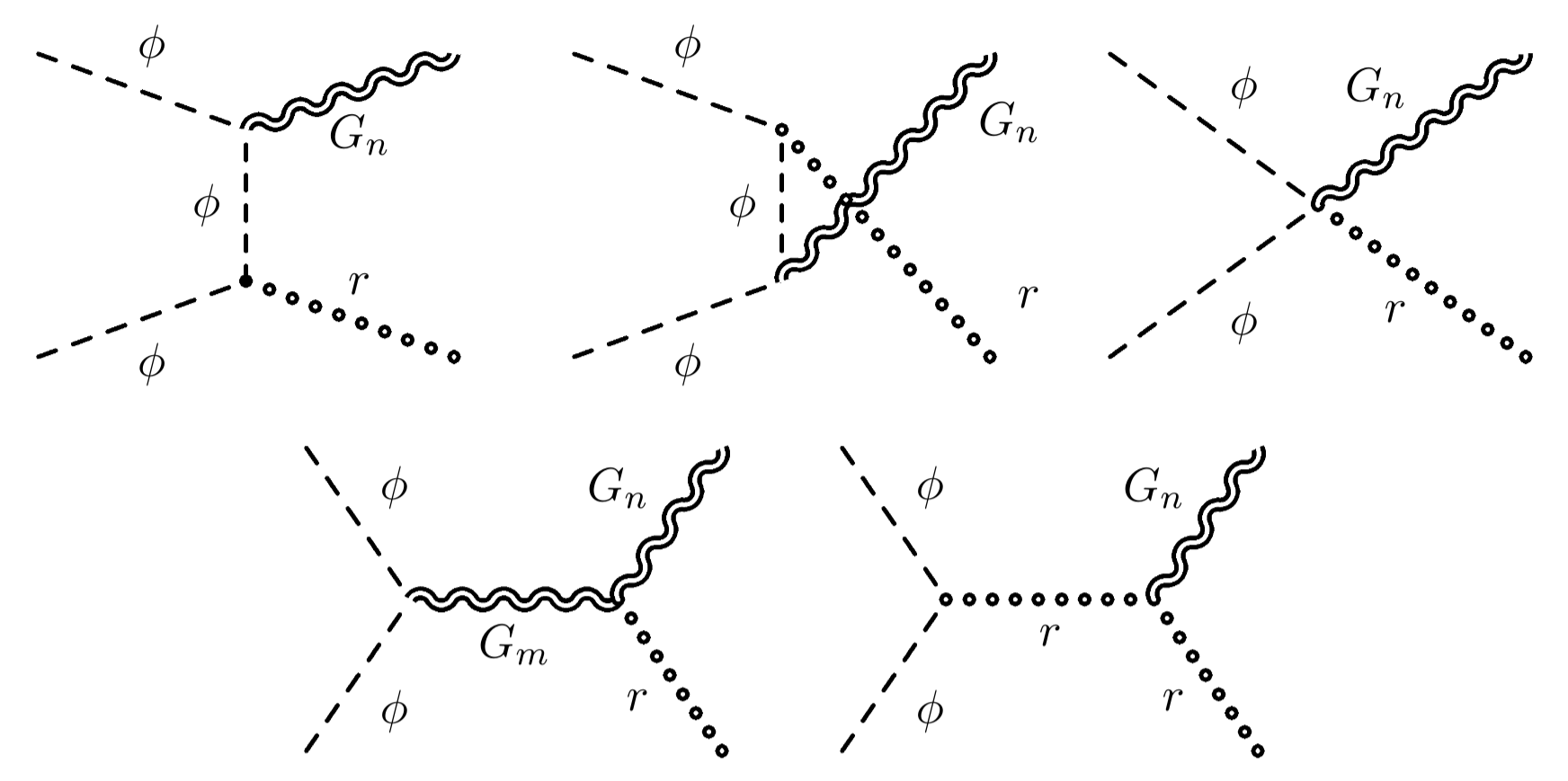}\par
\caption{$\phi\phi \longrightarrow r G_n$ | Representative diagrams contributing at leading order.}\label{radGravProd}
\end{figure}
As in the previous Section, we  expand $\mathcal{M}$ in $\sqrt{s}$ and report only  contributions that grow faster than $s$. Since a radion is produced, only one longitudinal graviton can be produced and the leading contribution to the matrix element grows as $\mathcal{O}(s^2)$. 
We find
\begin{align}
\mathcal{M}_0=\frac{i s^2 \left(\sum_m\frac{ k^2e^{-2\mu\pi}\tilde{\chi} _{\text{nmr}}}{ m_m^2}+\chi _{\text{nrr}}-1\right)}{24 \Lambda ^2 m_n^2} + \mathcal{O}(s)\,,
\end{align}
which leads to our only sum rule from radion final states:
\begin{equation}
    \sum\limits_m k^2e^{-2\mu\pi}\frac{\tilde{\chi} _{\text{nmr}}}{ m_m^2}=1-\chi _{\text{nrr}} \, .
\end{equation}
This condition relates the graviton-radion-radion coupling and the graviton-graviton-radion coupling.

\end{section}

\begin{section}{Sum rules \label{sec:sum_rules}}

After having identified the sum rules that ensure the cancellation of contributions that grow faster than $s$, we now show that these hold in the Randall-Sundrum model.
Before starting in earnest, it is important to note that the sum rules for the $\chi$-couplings can be derived from relations that only involve $\tilde{\chi}$ since their sums are related. In fact this even goes beyond the large $\mu$ limit and one can already show this at the level of the $a$ and $b$ coupling.   Integration by parts of $b_{nkm}$ leads to
\begin{equation}
    \begin{split}
        b_{nmk} & \equiv \int\limits_{-\pi}^\pi \diff{\varphi} \ A^4 \partial_\varphi \psi_n \partial_\varphi \psi_m \psi_k \overset{I.b.P.}{=}-\int\limits_{-\pi}^\pi \diff{\varphi} \  \psi_n \partial_\varphi\left(\partial_\varphi \psi_m  A^4\psi_k \right) = \\
        & = -\int\limits_{-\pi}^\pi\diff{\varphi} \psi_n \psi_k \partial_\varphi\left(A^4 \partial_\varphi\psi_m\right) -\int\limits_{-\pi}^\pi\diff{\varphi} \ A^4 \partial_\varphi \psi_k \partial_\varphi \psi_m \psi_n = (m_m r_c)^2 a_{nkm}-b_{mkn} \ .
    \end{split}
    \label{eq:ab_swap}
\end{equation}
The large $\mu$ limit we are interested in allows to separate the part that depends on $\mu$ from the rest. This was already achieved with the use of $\chi,\tilde{\chi}$ that are related to $a,b$ through
\begin{equation}
    \begin{split}
        a_{nmk}&=\chi_{nmk}\sqrt{\mu}e^{\mu\pi}\ ,\\
        b_{nmk}&=\tilde{\chi}_{nmk} \mu^{5/2}e^{-\mu\pi}\ .
    \end{split}
\end{equation}
Thus, the relation in eq.~\ref{eq:ab_swap} reduces to 
\begin{align}
    \gamma_k^2 \chi_{nmk}=\tilde{\chi}_{knm}+\tilde{\chi}_{kmn}\, .
    \label{eq:chichitilde_relation}
\end{align}
In the following we will first prove the sum rules analytically  before investigating their implications numerically.

\begin{subsection}{Analytical prove}

The relations we need for the cancellations
are summarized in Tab.~\ref{tab:sum_rules}.
In the following we will employ various properties of the Bessel functions, see for example \cite{abramowitz+stegun,Bessel_web}.
Note that it is easier to prove the first sum rule if the second one has already been established. Therefore, we will demonstrate that they hold in the order that makes the proves simpler and not in the order in which they appear in the matrix element expansion. 
\begin{table}[H]
\centering
\begin{tabular}{cc}
\toprule
Sum rule 1: & $\sum\limits_{m=1}^\infty \chi_{nkm}=1$\\
Sum rule 2: & $\sum\limits_{m=1}^\infty \chi_{nkm} \ \gamma_m^2 =\gamma_n^2+\gamma_k^2$ \\
Sum rule 3: &$\sum\limits_{m=1}^\infty \frac{\chi_{nkm}}{\gamma_m^2}(\gamma_n^2-\gamma_k^2)^2=(\gamma_n^2+\gamma_k^2)-6\tilde{\chi}_{nkr}\quad \overset{k=n}{\Longrightarrow}\quad  \tilde{\chi}_{nnr}=\frac{1}{3}\gamma_n^2$\\
Sum rule 4: &$\sum\limits_{m=1}^\infty \frac{\tilde{\chi}_{nmr}}{\gamma_m^2}=1-\chi_{nrr}$\\
\bottomrule
\end{tabular}
\caption{Sum rules needed for the cancellations.\label{tab:sum_rules}}
\label{eq:allsumrules}
\end{table}

Our starting point is the Fourier-Bessel expansion, see for example \cite{sneddon_1960}. If a function $f(x)$ is continuous on $[0, 1]$ such that $f(1)=0$, the integral
\begin{equation}
    \int\limits_0^1 \diff{x} \ x^{1/2} f(x)
\end{equation}
exists and is absolutely convergent, and $f(x)$ has limited total fluctuation, it can be expanded in series in terms of any Bessel-function $J_\nu$:
\begin{equation}
   f(x)= \sum\limits_{k=1}^\infty a_{\nu,k} J_\nu(\gamma_{\nu,k}x) \ ,
\end{equation}
where $\gamma_{\nu,k}$ is the $k-$th root of $J_\nu$ and
\begin{equation}
    a_{\nu,k}= \frac{2}{J_{\nu+1}(\gamma_{\nu,k})^2} \int\limits_0^1 \diff{u} \  u f(u) J_\nu(\gamma_{\nu,k} u) \ .
\end{equation}
Since we will work only with the roots of $J_1$, we define $\gamma_k$ as the $k-$th root of $J_1$ and we will set also $\nu=1$ such that
\begin{equation}
\label{fc-style}
    f(x)=2\sum\limits_{k=1}^\infty \frac{J_1(\gamma_k x)}{J_2(\gamma_k)^2}\int\limits_0^1 \diff{u} \  u f(u) J_1(\gamma_{k} u) \ .
\end{equation}
Let us recall that:
\begin{equation}
 \tilde{\chi}_{knm} \equiv -2\frac{\gamma_k \gamma_n}{J_0(\gamma_k)J_0(\gamma_n)J_0(\gamma_m)}\int\limits_0^1 \diff{u} \ u^3 J_1(\gamma_k u)J_1(\gamma_n u)J_2(\gamma_m u) \ .
\end{equation}
A convenient choice of $f(x)$ that satisfies $f(1)=0$ is given by:
\begin{equation}
\label{eq:rep1}
    f(x)=\frac{\gamma_n}{J_0(\gamma_n)J_0(\gamma_m)}J_1(\gamma_n x)J_2(\gamma_m x) x^2 \ .
\end{equation}
Using the fact that $J_0(\gamma_n)=-J_2(\gamma_n)$ and multiplying and dividing by $\gamma_k$, we have:
\begin{equation}
\label{eq:rep2}
    f(x)=\sum\limits_{k=1}^\infty \frac{J_1(\gamma_k x)}{\gamma_k J_2(\gamma_k)}\tilde{\chi}_{knm} \ .
\end{equation}
Differentiating both sides:
\begin{equation}
    f'(x)=\frac{1}{2}\sum\limits_k \tilde{\chi}_{knm}\left[\frac{J_0(\gamma_kx)-J_2(\gamma_kx)}{J_2(\gamma_k)}\right]\overset{x=1}{\rightarrow}-\sum\limits_k \tilde{\chi}_{knm}  \ .
\end{equation}
But on the other side:
\begin{equation}
    f'(x)|_{x=1}=-\gamma_n^2 \ .
\end{equation}
We then have a very helpful relation that will be useful soon:
\begin{equation}
    \label{eq:mainsumrule}
    \sum\limits_k \tilde{\chi}_{knm} =\gamma_n^2 \ .
\end{equation}
Combining this expression and eq.~\ref{eq:chichitilde_relation} directly leads to
\begin{equation}
      \sum\limits_k \chi_{nmk} \gamma_k^2 =(\gamma_n^2+\gamma_m^2)
\end{equation}
thus proving {\bf sum rule 2}.
By making further use of eq. \ref{eq:chichitilde_relation} we can  derive other sum rules. Permuting the indices we have two equations:
\begin{equation}
    \tilde{\chi}_{nmk}=\gamma_m^2 \chi_{nkm}-\tilde{\chi}_{mkn} \quad , \quad  \tilde{\chi}_{nkm}=\gamma_k^2 \chi_{nkm}-\tilde{\chi}_{mkn} \ .
\end{equation}
Using the symmetry properties of $\chi_{\Vec{n}}$ and $\tilde{\chi}_{\Vec{n}}$:
\begin{equation}
\begin{split}
        \frac{\tilde{\chi}_{nmk}+\tilde{\chi}_{mkn}}{\gamma_m^2} & =\frac{\tilde{\chi}_{nkm}+\tilde{\chi}_{mkn}}{\gamma_k^2} \ , \\
        \Longrightarrow \quad \tilde{\chi}_{mkn}\left(\frac{1}{\gamma_m^2}-\frac{1}{\gamma_k^2} \right) & = \frac{\tilde{\chi}_{nkm}}{\gamma_k^2}-\frac{\tilde{\chi}_{nmk}}{\gamma_m^2} \ ,\\
        \Longrightarrow \quad  \sum\limits_{n=1}^\infty \tilde{\chi}_{mkn}\left(\frac{1}{\gamma_m^2}-\frac{1}{\gamma_k^2} \right) &= \sum\limits_n \frac{\tilde{\chi}_{nkm}}{\gamma_k^2}-\sum\limits_n\frac{\tilde{\chi}_{nmk}}{\gamma_m^2} \overset{Eq.\ref{eq:mainsumrule}}{=} 0 \ .
\end{split}
\end{equation}
This implies that:
\begin{equation}
    \sum\limits_{k=1}^\infty \tilde{\chi}_{nmk} = 0 \quad \text{if} \quad n\neq m \ .
\end{equation}
At this point we have finally the last relation for the case $n\neq m$. Using again eq.~\ref{eq:chichitilde_relation} we find
\begin{equation}
    \gamma_k^2 \sum\limits_n \chi_{nkm} =\sum\limits_n \tilde{\chi}_{mkn} + \sum\limits_n \tilde{\chi}_{nkm} = 0 + \gamma_k^2\ ,
\end{equation}
which proves {\bf sum rule 1} for $n\neq m$.
For the case with $n=m$ we need to work a little harder. 
The starting point is eq. \ref{eq:rep2} combined with the definition of $f(x)$ shown in eq. \ref{eq:rep1}. 
Multiplying both sides by $x^2$ and integrating from 0 to 1 we get rid of the extra Bessel functions in the sum:
\begin{equation}
\label{eq:rep3}
    \int\limits_0^1 \diff{x} \ x^2 f(x) = \sum\limits_k \frac{\tilde{\chi}_{knm}}{\gamma_k^2 } \ .
\end{equation}
In the special case $n=m$ the integral can be performed analytically  
\begin{equation}
    \int\limits_0^1 \diff{x}\ x^4 J_1(\gamma_n x) J_2(\gamma_n x)= \frac{J_2(\gamma_n)^2}{2\gamma_n}\ .
\end{equation}
Including the coefficients of $f(x)$, it follows that
\begin{equation}
     \sum\limits_k \frac{\tilde{\chi}_{knn}}{\gamma_k^2} =\frac{1}{2}\,.
\end{equation}
Using again Equation \ref{eq:chichitilde_relation} and setting $n=m$ we get
\begin{equation}
    \gamma_k^2 \chi_{nnk}=2\tilde{\chi}_{knn}  \quad \Longrightarrow \quad    \sum\limits_k \chi_{nnk}= 2\sum\limits_k \frac{\tilde{\chi}_{knn}}{\gamma_k^2}=  1\ ,
\end{equation}
which proves {\bf sum rule 1} for $m=n$.

Next we tackle the relation involving $\chi_{nrr}$. 
Let us recall the definition of the radion coefficients:
\begin{equation}
    \begin{split}
        &\tilde{\chi}_{knr}\equiv 2\frac{\gamma_k \gamma_n}{J_0(\gamma_n)J_0(\gamma_m)}\int\limits_0^1 \diff{u} \ u^3 J_1(\gamma_k u)J_1(\gamma_n u) \ ,\\
        &\chi_{nrr}\equiv -2\frac{1}{J_0(\gamma_n)}\int\limits_0^1 \diff{u} \ u^3 J_2(\gamma_n u) = -\frac{2}{\gamma_n}\frac{J_3(\gamma_n)}{J_0(\gamma_n)} \ .
    \end{split}
\end{equation}
As before, the way to get the desired result is a good choice of $f(x)$; in this case we use:
\begin{equation}
    f(x) = x^2 \frac{J_1(\gamma_n x)}{J_0(\gamma_n x)}\gamma_n \ ,
\end{equation}
from which follows
\begin{equation}
    f(x)=\sum\limits_k \frac{J_1(\gamma_k x)}{\gamma_k J_0(\gamma_k)}\tilde{\chi}_{knr} \ .
\end{equation}
Multiplying both sides by $x^2$ and integrating from 0 to 1 we find
\begin{equation}
    \sum\limits_k \frac{\tilde{\chi}_{knr}}{\gamma_k^2}=-\frac{\gamma_n}{J_0(\gamma_n)}\int\limits_0^1 \diff{x}\ x^4 J_1(\gamma_n x)=-\frac{4 J_3(\gamma_n )-\gamma_n  J_4(\gamma_n )}{\gamma_n  J_0(\gamma_n )} \ .
\end{equation}
Using the properties of the Bessel functions, this last result can be written as:
\begin{equation}
   -\frac{\gamma_n}{J_0(\gamma_n)}\int\limits_0^1 \diff{x}\ x^4 J_1(\gamma_n x)= 1-\chi_{nrr} \ .
\end{equation}
Then the desired result follows
\begin{equation}
    \sum\limits_k \frac{\tilde{\chi}_{knr}}{\gamma_k^2}=1-\chi_{nrr}\ ,
\end{equation}
which proves {\bf sum rule 4}.
Now we turn to sum rule 3, which will turn out to be the most complicated one. With eq.~\ref{eq:chichitilde_relation} we get
\begin{equation}
    \gamma_n^2 \chi_{nmk}=\tilde{\chi}_{nmk}+\tilde{\chi}_{nkm} \ \ , \ \  \gamma_m^2 \chi_{nmk}=\tilde{\chi}_{mnk}+\tilde{\chi}_{mkn} \\ \Longrightarrow \\ (\gamma_n^2-\gamma_m^2)\chi_{nmk}=\tilde{\chi}_{nkm}-\tilde{\chi}_{kmn} \ .
\end{equation}
Multiplying both sides by $(\gamma_n^2-\gamma_m^2)$, dividing by $\gamma_k^2$ and summing over $k$ we get
\begin{equation}
    \sum\limits_k \chi_{nmk}\frac{(\gamma_n^2-\gamma_m^2)^2}{\gamma_k^2}=(\gamma_n^2-\gamma_m^2)\sum\limits_k \frac{\tilde{\chi}_{knm}-\tilde{\chi}_{kmn}}{\gamma_k^2} \ .
\end{equation}
To shorten the notation, let us define the following quantity:
\begin{equation}
    f_{nm}\equiv \sum\limits_k\frac{\tilde{\chi}_{knm}}{\gamma_k^2} \quad \Longrightarrow \quad \sum\limits_k \chi_{nmk}\frac{(\gamma_n^2-\gamma_m^2)^2}{\gamma_k^2}=(\gamma_n^2-\gamma_m^2)(f_{nm}-f_{mn}) \ .
\end{equation}
We proceed by using the result of eq. \ref{eq:rep3}.
One finds
\begin{equation}
   f_{nm}\equiv \sum\limits_k \frac{\tilde{\chi}_{knm}}{\gamma_k^2}= \mathcal{N}\gamma_n\int\limits_0^1 \diff{x} \ x^4 J_1(\gamma_n x) J_2(\gamma_m x) \ ,
\end{equation}
where we have defined a normalization factor
\begin{equation}
    \mathcal{N}\equiv \frac{1}{J_0(\gamma_n)J_0(\gamma_m)}
\end{equation}
to get more compact expressions.
We can express this integral in terms of $\tilde{\chi}_{nmr}$, which is the quantity we are interested in. To see this connection we need the following two relations. 
\subparagraph{One: $\gamma_m^2 f_{nm}+\gamma_n^2f_{mn}=3\tilde{\chi}_{nmr}$} For this we exploit the following property of the Bessel functions:
\begin{equation}
\label{eq:rel1}
   J_2(\alpha x)=\frac{1}{\alpha}\left(\frac{1}{x}J_1(\alpha x)-\partial_x(J_1(\alpha x))\right) \quad ,\quad \forall \alpha \in \mathbb{C} \ .
\end{equation}
Then, it follows that:
\begin{equation}
    \begin{split}
         f_{nm}&=\gamma_n \mathcal{N}\int\limits_0^1 \diff{x} \ x^4J_1(\gamma_n x) J_2(\gamma_m x)\\& \overset{Eq.\ref{eq:rel1}} {=} \gamma_n \mathcal{N}\int\limits_0^1 \diff{x} \ x^4J_1(\gamma_n x) \frac{1}{\gamma_m}\left(\frac{1}{x}J_1(\gamma_m x)-\partial_x(J_1(\gamma_m x))\right)  \\
         &= \frac{1}{2\gamma_m^2}\tilde{\chi}_{nmr}-\frac{\gamma_n}{\gamma_m}\mathcal{N}\int\limits_0^1 \diff{x} \ x^4J_1(\gamma_n x) \partial_x(J_1(\gamma_m x))\\ & \overset{I.b.P.}{=}\frac{1}{2\gamma_m^2}\tilde{\chi}_{nmr}+\frac{\gamma_n}{\gamma_m}\mathcal{N}\int\limits_0^1 \diff{x} \ \partial_x(x^4J_1(\gamma_n x)) J_1(\gamma_m x) = \\
         & \overset{I.b.P.+ Eq.\ref{eq:rel1}}{=} 3 \frac{\tilde{\chi}_{nmr}}{\gamma_m^2}-\frac{1}{\gamma_m^2}f_{mn} \ ,
    \end{split}
\end{equation}
    which proves the statement above.
\subparagraph{Two: $ f_{nm}+f_{mn}=1$}
This is simpler and it relies on the following relation:
\begin{equation}
\label{eq:rel2}
    x^2 J_1(\alpha x)=\partial_x \left(\frac{1}{\alpha}x^2 J_2(\alpha x)\right) \quad ,\quad \forall \alpha \in \mathbb{C} \ .
\end{equation}
It follows that:
\begin{equation}
    \begin{split}
        f_{nm}&=\gamma_n \mathcal{N}\int\limits_0^1 \diff{x} \ x^4J_1(\gamma_n x) J_2(\gamma_m x) \overset{Eq.\ref{eq:rel2}}{=}  \mathcal{N}\int\limits_0^1 \diff{x} \ x^2 \partial_x \left(x^2 J_2(\gamma_n x)\right) J_2(\gamma_m x)= \\
        &\overset{I.b.P.}{=} 1 -\mathcal{N}\int\limits_0^1 \diff{x} \ x^2 J_2(\gamma_n x) \partial_x \left(x^2 J_2(\gamma_m x)\right) \overset{Eq.\ref{eq:rel2}}{=} 1-f_{mn}
    \end{split}
\end{equation}
and hence
\begin{equation}
\label{eq:2}
    f_{nm}+f_{mn}=1 \ .
\end{equation}

Using the previous two relations it follows that:
\begin{equation}
    \begin{split}
        f_{nm}-f_{mn}\overset{Eq.\ref{eq:2}}{=}-1+2f_{nm}\ .
    \end{split}
\end{equation}
Combining relation one from above and Equation \ref{eq:2}:
\begin{equation}
    \begin{split}
        f_{nm}=\frac{\gamma_n^2-3\tilde{\chi}_{nmr}}{\gamma_n^2-\gamma_m^2} \ .
    \end{split}
\end{equation}
Then the final result follows:
\begin{equation}
    \begin{split}
        \sum\limits_k \chi_{nmk}\frac{(\gamma_n^2-\gamma_m^2)^2}{\gamma_k^2}&=(\gamma_n^2-\gamma_m^2)\sum\limits_k \frac{\tilde{\chi}_{knm}-\tilde{\chi}_{kmn}}{\gamma_k^2}=(\gamma_n^2-\gamma_m^2)(f_{nm}-f_{mn})=\\
        &=(\gamma_n^2+\gamma_m^2)-6\tilde{\chi}_{nmr}\ ,
    \end{split}
\end{equation}
which proves {\bf sum rule 3}. Thus we have demonstrated that the sum rules required to cancel the contributions that grow faster than $s$ are fulfilled in the Randall-Sundrum model in the large $\mu$ limit. 

A few comments about the limitations of our analysis are in order. First, we only considered the large $\mu$ limit. In principle, we expect the cancellation to work even without this restriction. However, a possible generalization to this case comes with a number of complications. The zero-mode of the graviton, which does not contribute in the limit we considered, can no longer be neglected. This can be accommodated with minimal changes if we switch the discussion from $\chi$ and $\tilde{\chi}$ to $a$ and $b$. However, we cannot employ the limiting form of the wave-functions and masses. Therefore, relations that directly rely on the properties of the wave-functions and their Fourier-Bessel expansion need to be generalized. Second, it is also worth pointing out that the radion is massless in our analysis. A massless scalar that with a coupling stronger than gravity would spoil General Relativity. Therefore, this is not acceptable if the model is supposed to incorporate the world we live in. The problem can be solved by a radion mass. This can be achieved by a mechanism that stabilizes the radius of the extra-dimension; a concrete example is the Goldberger-Wise mechanism  \cite{Goldberger:1999uk}. It relies on a new bulk scalar that mixes with the radion and, as a consequence, the single massless radion gets replaces by a KK-tower of massive scalar fields \cite{Csaki:2000zn}. An analysis of the impact of the Golberger-Wise mechanism on the unitarization of the matrix elements goes beyond the scope of this work. However, we expect that the basic conclusions will remain the same. To test this assumption we have added a radion mass $m_r$ by hand and find that the sum rules remain unaffected for $m_r\ll \sqrt{s}$.

\end{subsection}

\begin{subsection}{Numerical study}
  \begin{figure}[tbh] 
\centering%
\subfigure[{}\label{fig:011}]%
{\includegraphics[width=0.32\textwidth]{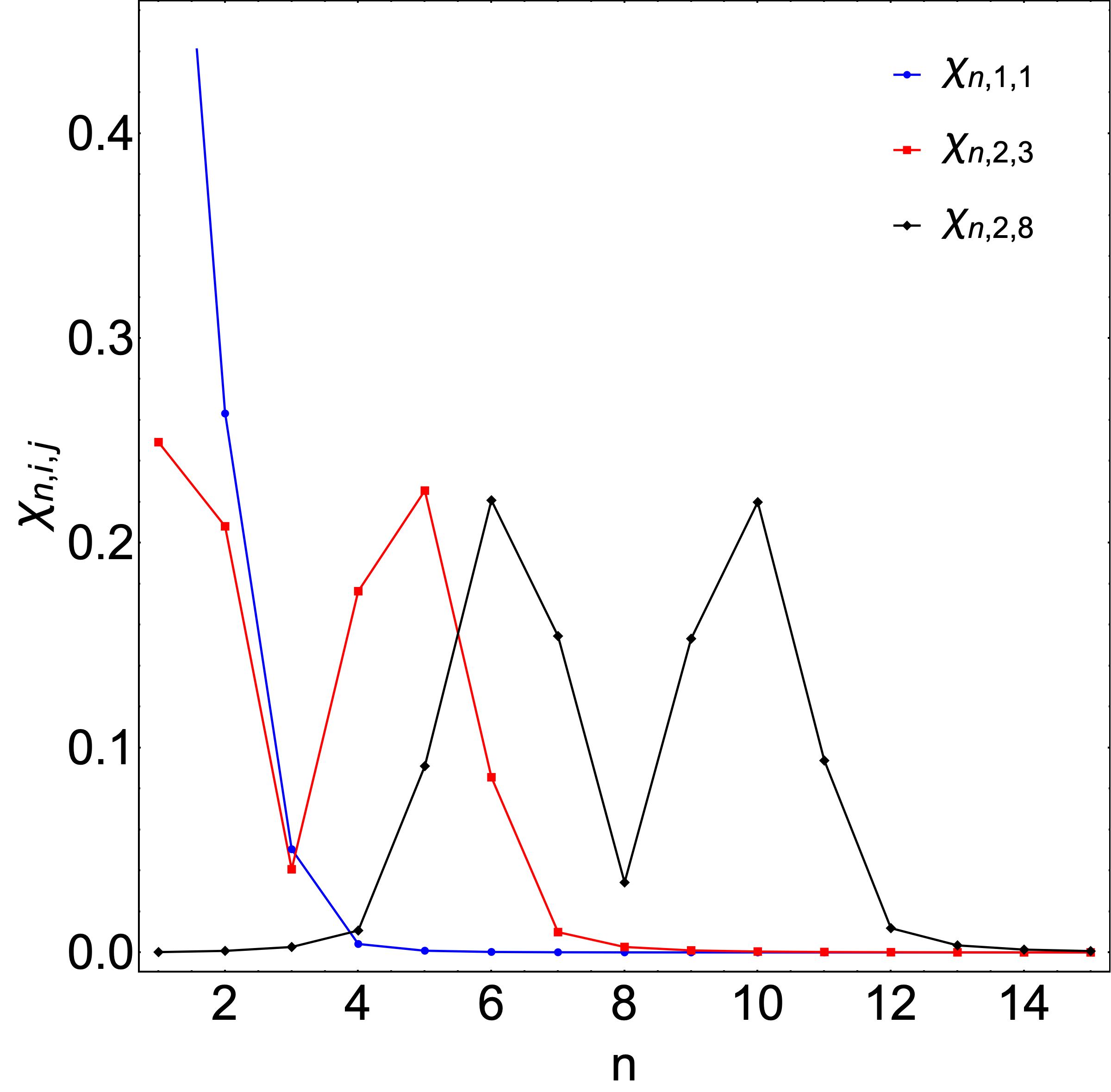}}\
\subfigure[{}\label{fig:022}]%
{\includegraphics[width=0.32\textwidth]{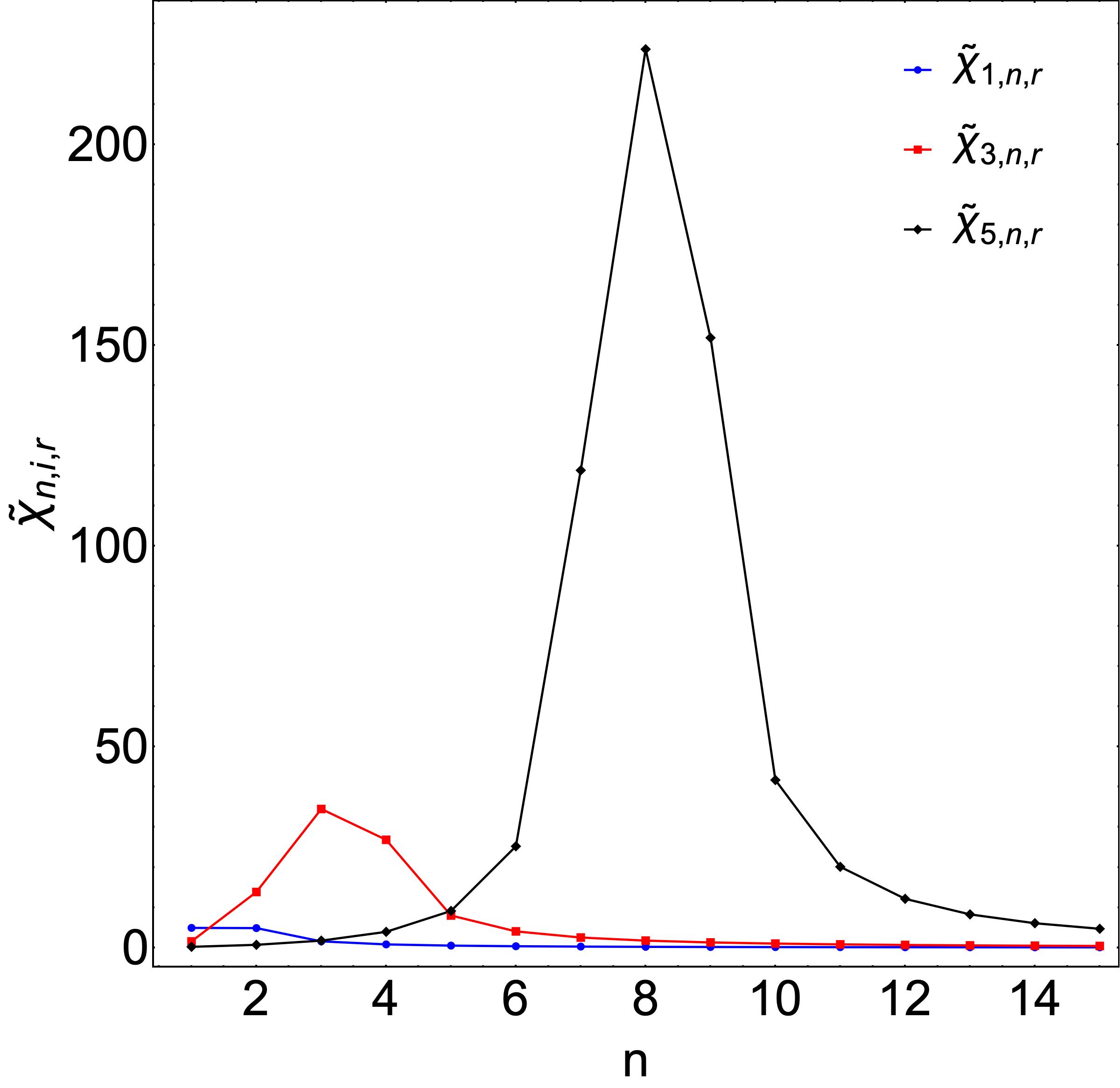}}
\
\subfigure[{}\label{fig:033}]%
{\includegraphics[width=0.335\textwidth]{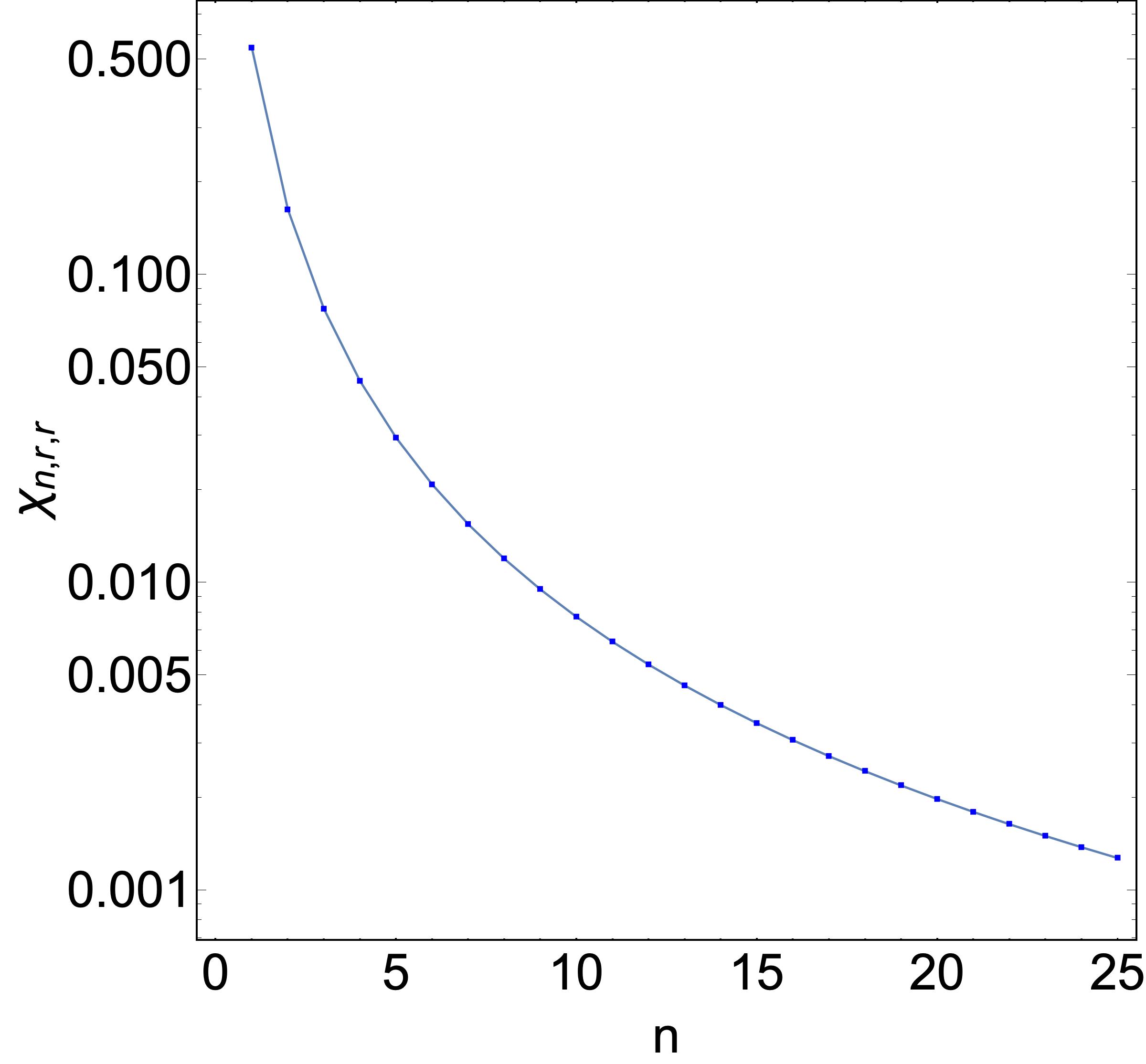}}
\caption{Numerical values of the relevant coefficients for a  representative set of gravitons.}
\label{fig:couplings}
\end{figure}

 \begin{figure}[tbh] 
\centering%
\subfigure
{\includegraphics[width=0.44\textwidth]{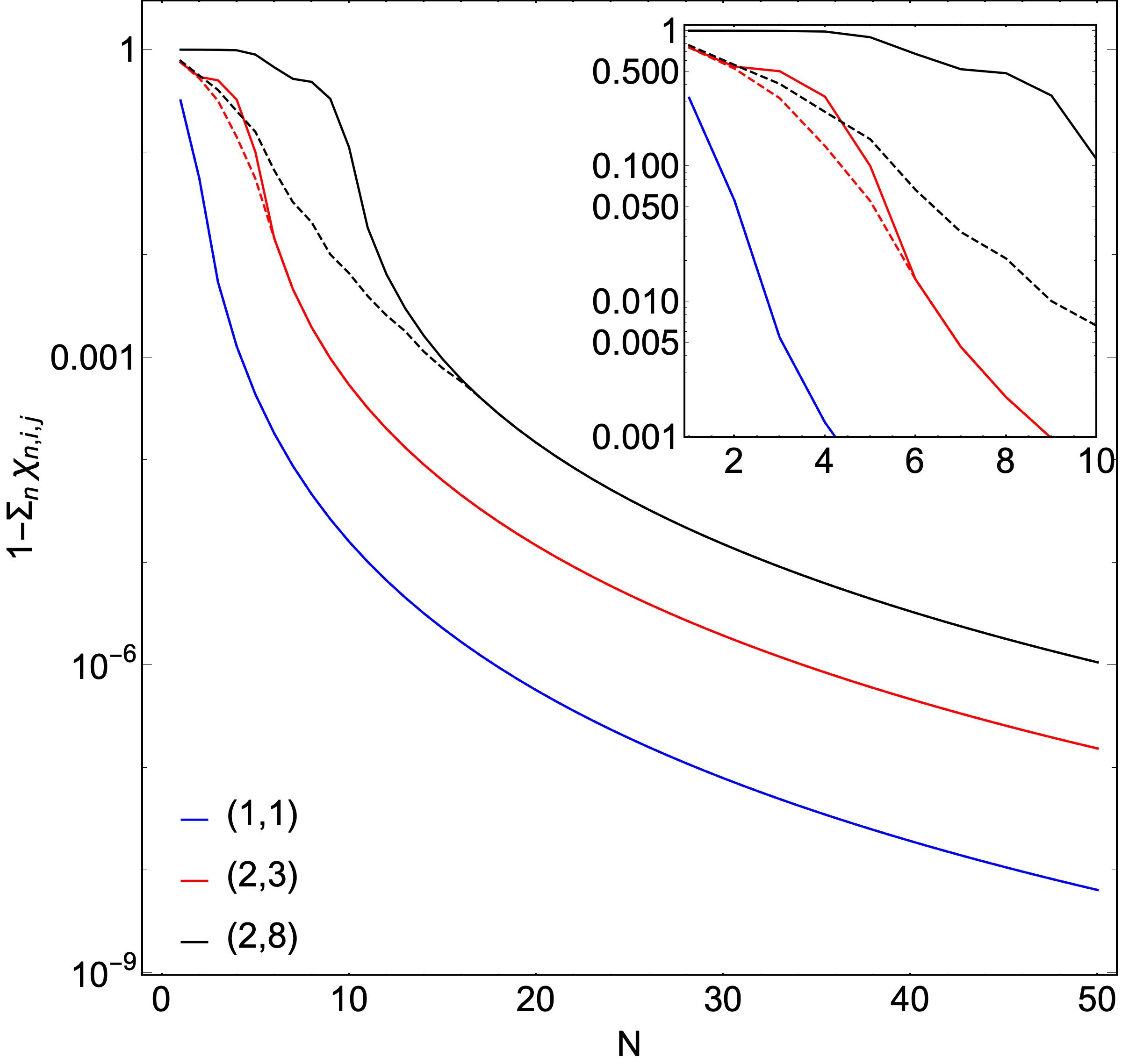}}\
\subfigure
{\includegraphics[width=0.44\textwidth]{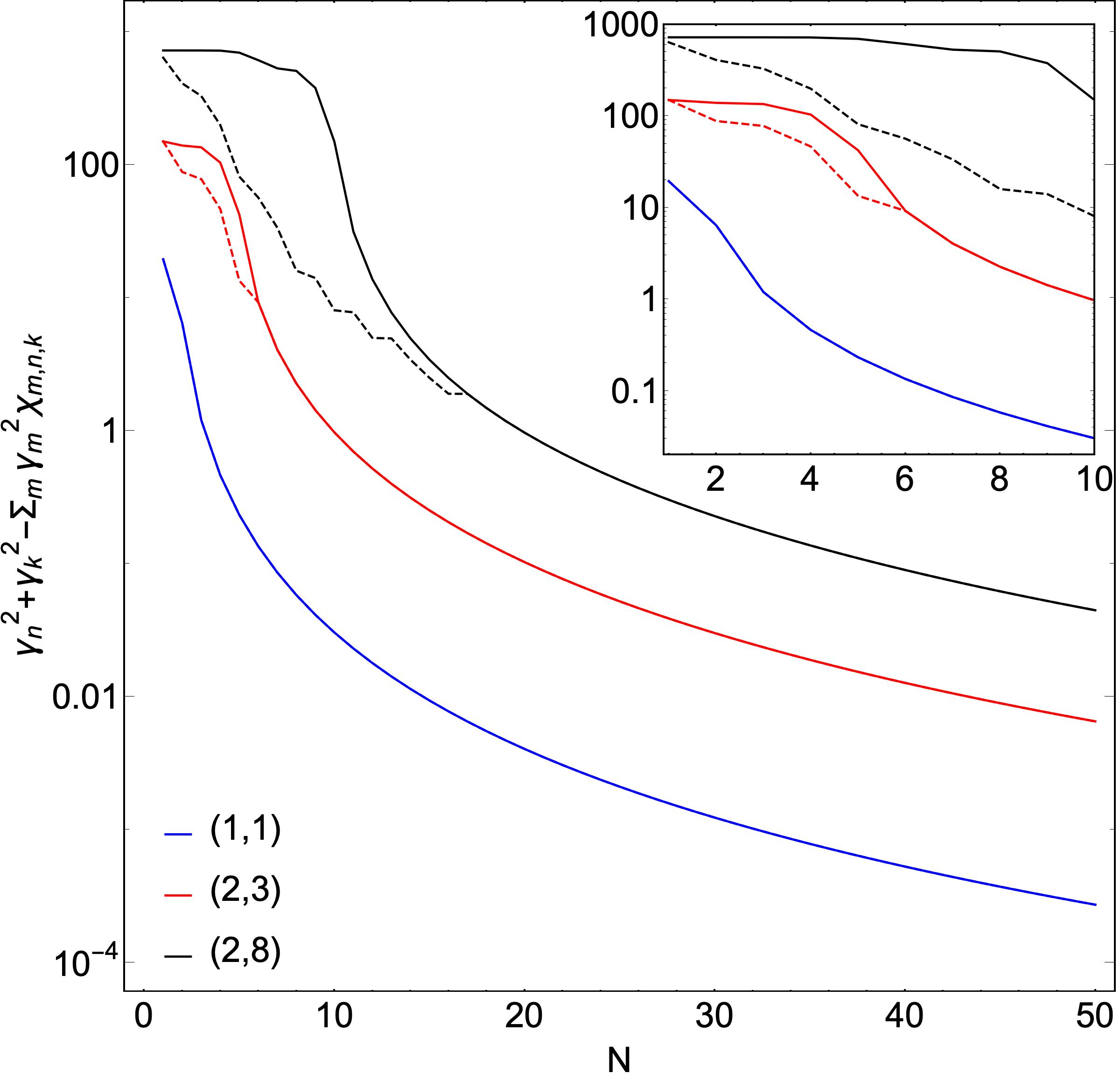}}
\caption{Illustration of the first (left) and second (right) sum rule for a truncated KK-tower as a function of the number of included gravitons N. For each sum rule we show three representative choices of the graviton final state as indicated by the inset in the panels. We show two prescriptions for the order in which gravitons are added. The solid lines show a standard truncation where all gravitons up to the $N$th are added while the dashed line corresponds to in improved prescription where the gravitons are added by the size of $\chi_{knm}$.  }
\label{fig:numerical-sum-rules1}
\end{figure}

 \begin{figure}[tbh] 
\centering%
\subfigure
{\includegraphics[width=0.45\textwidth]{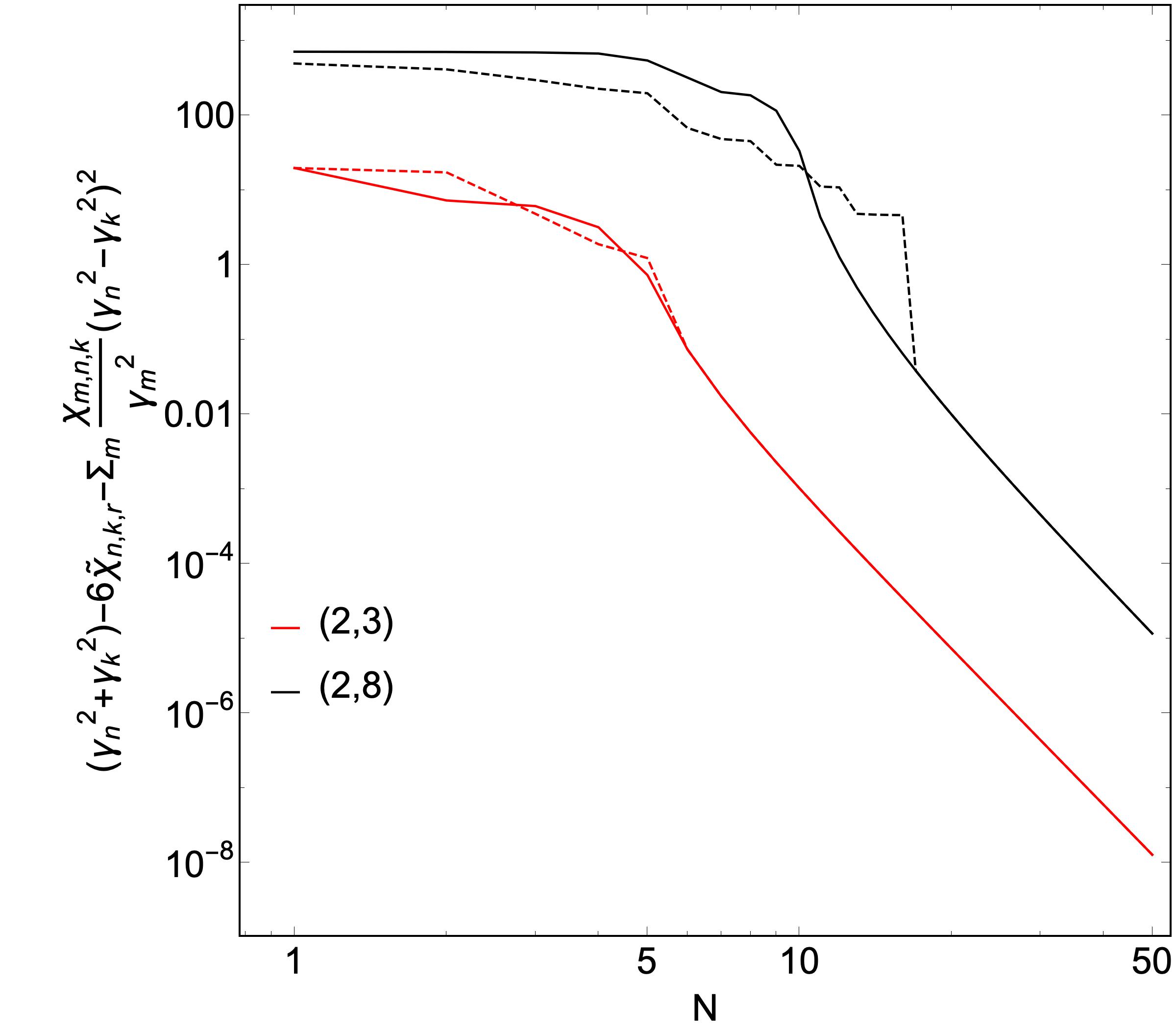}}
\
\subfigure
{\includegraphics[width=0.44\textwidth]{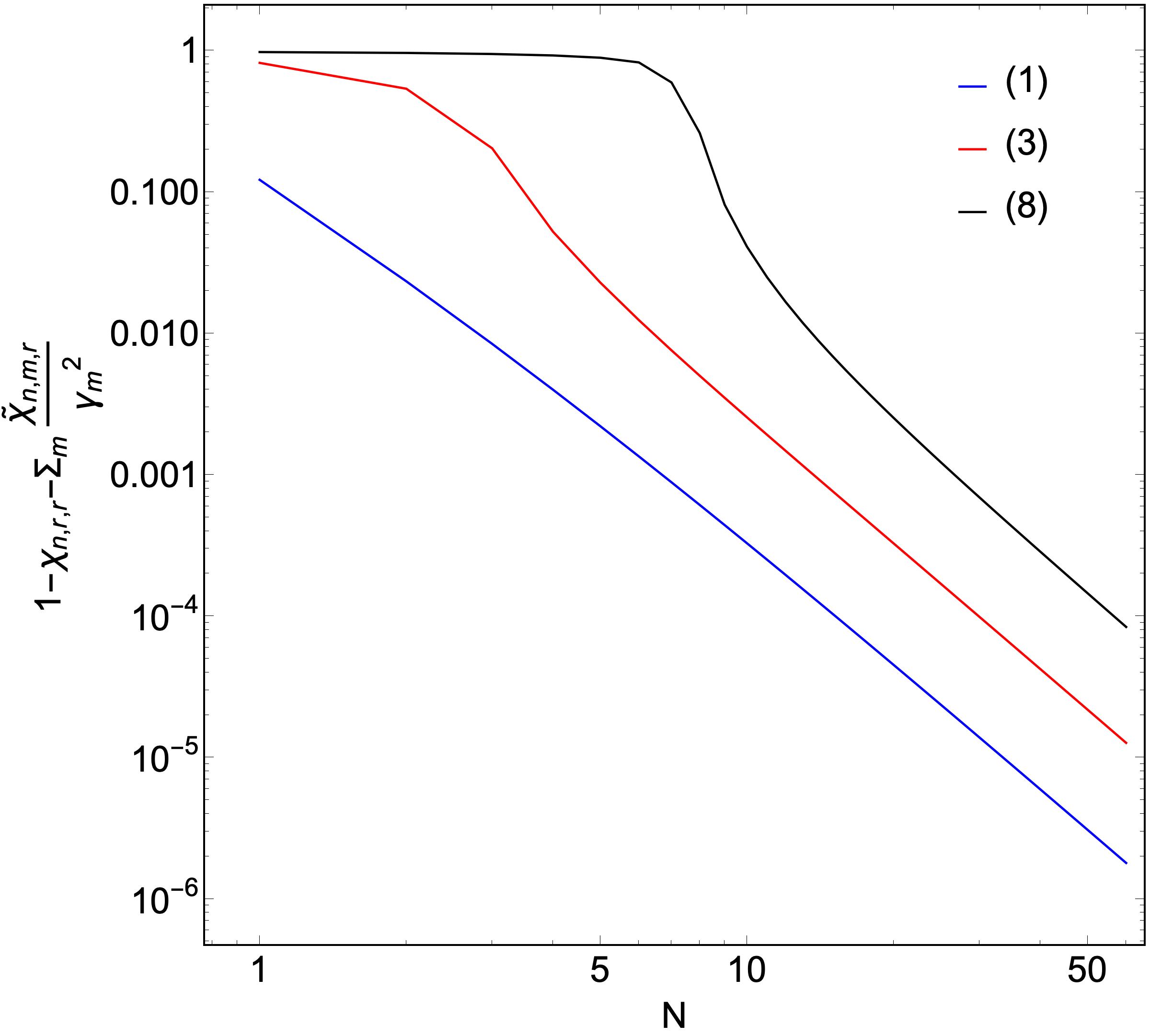}}
\caption{Illustration of the third (left) and forth (right) sum rule for a truncated KK-tower as a function of the number of included gravitons N. For the third sum rule we show two ways of adding the gravitons as in Fig.~\ref{fig:numerical-sum-rules1}. }
\label{fig:numerical-sum-rules2}
\end{figure}
 We have shown that the sum rules are fulfilled analytically but the question of how fast they converge remains.  In particular if one is interested in a quantitative study outside of the high energy limit it is often not feasible to impose the sum rules directly on the matrix element and one might want to work with a truncated KK-tower. Therefore, it is of great interest to investigate how many gravitons need to be summed to get a meaningful numerical result. 

Let us start by looking at the coefficients  $\chi_{nkm}$, $\tilde{\chi}_{nkr}$ and $\chi_{nrr}$. We show their values for a few representative combinations of gravitons in Fig.~\ref{fig:couplings}. Looking at the $\chi_{n_1,n_2,m}$ (left panel) we observe that they are relatively small with the largest in the ballpark of $0.20-0.25$. The distribution is bi-modal with  peaks around $m=|n_1\pm n_2|$ and falls of quite fast when this is not fulfilled. This is an important observation that allows us to optimize the number of gravitons we need to include in order to get a good approximation of the results.
The situation is somewhat similar in the case of $\tilde{\chi}_{nkr}$. Here, the couplings are largest when $n=k$ and fall off quite fast away from this. We observe a growth of $\tilde{\chi}_{nnr}$ with $n$ which is expected since we found analytically that  $\tilde{\chi}_{nnr}\propto \gamma^2_n$. However, this growth has no implication for sum rule 4 since only the combination $\tilde{\chi}_{nmr}/\gamma^2_m$ enters. The third type of coefficients is less interesting than the others. As can be seen in the third panel, $\chi_{nrr}$  decreases rather fast as $n$ increases and has no notable features.

Now we turn to study the sum rules in the case of a truncated KK-tower. In Fig.~\ref{fig:numerical-sum-rules1} and Fig.~\ref{fig:numerical-sum-rules2} we show the difference of the sum rule from zero as a function of the number of gravitons in the sum for some representative benchmarks. For light gravitons the convergence is relatively fast and considering the first ten massive KK-modes leads to a reduction of the numerical coefficient by a factor $\leq 10^{-3}$ for both the first and the second sum rule. However, if we consider heavier gravitons the situation is a bit different. Since the $\chi_{n_1 n_2 m}$ coefficients peak at around $m=|n_1\pm n_2|$  adding light graviton does not improve the cancellation significantly until $N\gtrsim n_1+n_2$. This can be partially ameliorated if we do not add the gravitons by mass but by largest $\chi$ factor. This trick works best for sum rules 1 and 2  and speeds up the cancellation for moderate values of $N$ while the large $N$ behavior is essentially identical. However, even in this case the precision to which the sum rules are fulfilled lags behind the one achieved for the light spin-2 fields. In the case of sum rule 3, it does not help much and can even inhibit the cancellation or certain choices of $N$ as can be seen in Fig.~\ref{fig:numerical-sum-rules2}. For two identical gravitons in the final state sum rule 3 reduces to an analytical expression for $\chi_{nnr}$. As it does not depend on the third graviton any more the sum is trivial and the level of precision and any deviation from zero is just related to the numerical precision of the evaluation of $\chi_{nnr}$. Therefore, we do not show it in the figure.

The level of precision to which the high energy growth has to cancel in a numerical study will depend on the computation and the desired precision. However, it appears that just adding one or two gravitons is not enough for most applications. Even for the pair production of the lightest graviton the first five modes need to be considered if a suppression of the unphysical contributions by a factor of $\approx 10^{-3}$ is desired.

\end{subsection}

\end{section}

\begin{section}{Conclusions}
\label{sec:conclusions}

Extra-dimensional theories have received considerable interest in recent years. They provide attractive models for physics beyond the Standard Model that can help address unsolved problems of high energy physics such as the hierarchy problem, dark matter, or the structure of the SM Yukawa couplings.
Higher dimensional theories are interesting for other reasons as well. They allow for intriguing modifications of gravity and give rise to massive spin-2 fields. Despite many years of study, massive gravity is still a topic of current research and far from being completely understood. Extra-dimensions provide test cases for more general theories and can help out in our understanding of massive spin-2 fields.

We choose a particular realization of warped extra-dimensions for our study. To be concrete, we focused on a simplified version of the well-known Randall-Sundrum model with a toy matter sector consisting of a scalar with gravitational interactions. The structure of the matrix elements for KK-graviton production from pairs of scalars in the initial state are very intricate and we find that individual contributions grow as fast as $s^3$ in the high energy limit.
Taken at face value, this points towards a breakdown of perturbative unitarity well below the fundamental scale of high dimensional gravity.
However, a closer study reveals that the theory enforces correlations between the graviton (or radion) self-interactions which conspire to cancel terms that grow faster than $s$. These correlations can be interpreted as sum rules for the coefficients of the interactions between gravitational fields. We proved analytically that they hold in the large $\mu$ limit of the Randall-Sundrum model. Thus the validity of the theory is restored once the full KK-tower is included in the calculations. Similar results have been obtained in studies of KK-graviton scattering both in warped extra-dimensions and other geometries \cite{Chivukula:2020hvi,Bonifacio:2019ioc}. These works found sum rules that relate the coefficients of the gravitation three-point interaction to the four point vertices.  We find a separate set of sum rules that connect the graviton and radion interactions to the coupling with matter fields thus complementing earlier work. This is interesting from a  purely theoretical perspective since it adds a  second set of conditions for a realistic theory. 
In addition, our results are also relevant to phenomenological studies. For example, we expect effects in the production of gravitationally interacting dark matter that we intend to study in a forthcoming publication.

\newpage
\section*{Acknowledgments}
In our calculations we used \textit{xAct} \cite{xAct} and \textit{xPert} \cite{xPert} for the expansion of the Lagrangian,
\textit{FeynRules} \cite{Christensen_2009} to derive the Feynman rules and  \textit{FeynCalc} \cite{MERTIG1991345,Shtabovenko_2016,Shtabovenko_2020} for symbolic manipulation of the matrix elements. After this work appeared on the arXiv we received communications by R. Sekhar Chivukula on behalf of the authors of \cite{Chivukula:2020hvi}. They are currently working on a related study and confirm our main results summarized in Tab.~\ref{tab:sum_rules}.
\end{section}

\FloatBarrier


\begin{appendices}
\numberwithin{equation}{section}

\section{Lagrangian Expansion\label{Appx:Interactions}}
We present the results of the weak-field expansion of the metric 
\begin{equation}
\label{perturbedmetric}
    ds^2 = A(z)^2\left(e^{-2\hat{u}}(\eta_{\mu\nu}+\kappa \ \hat{h}_{\mu\nu})\diff{x}^\mu \diff{x}^\nu -(1+2\hat{u})^2 \diff{z}^2\right) \ ,
\end{equation}
where the coordinate $z$ is related to $y$ through
\begin{equation}
\label{transofrmations}
\text{d}z=A(y)^{-1}\text{d}y \quad,\quad\pder{}{z}=A(y)\pder{}{y} 
\end{equation}
and where we have defined $\kappa \equiv \frac{2}{M_5^{3/2}}$ and $\hat{u}$ as
\begin{equation}
    \hat{u}(x,y) \equiv \kappa \ \frac{\hat{r}(x)}{2\sqrt{6}} \ e^{2k|y|} \ .
\end{equation}
We also adopt the notation $\partial_\mu \equiv _{,\mu}$ and $\partial_z \equiv \ '$. In this coordinate, eq. \ref{eq:mass-phi} reads
\begin{equation}
    \der{}{z}\left[A(z)^3\der{\psi_n}{z}\right]=A^2(3A'\psi_n'+A\psi_n'')=-m_n^2 A^3 \psi_n \ .
\end{equation}
For completeness, we first report the volume elements of the bulk
\begin{equation}
\label{detBU}
\begin{split}
 \sqrt{G}=& A^5\left[ 1+\kappa  \left(\frac{1}{2} \hat{h}-\frac{\hat{r}}{\sqrt{6} A^2}\right)+\kappa ^2 \left(-\frac{\hat{r} \hat{h}}{2 \sqrt{6} A^2}-\frac{1}{4} \hat{h}^{\lambda\mu} \hat{h}_{\mu\lambda} +\frac{1}{8} \hat{h}^2\right)+ \right. \\
 &\left. + \kappa ^3 \left(\frac{\hat{r} \hat{h}^{\lambda\mu} \hat{h}_{\mu\lambda}}{4 \sqrt{6} A^2}-\frac{\hat{h} ^2\hat{r}}{8 \sqrt{6} A^2}+\frac{1}{6} \hat{h}^{\lambda\mu} \hat{h}_{\nu\lambda} \hat{h}_\mu^\nu -\frac{1}{8} \hat{h} \hat{h}^{\mu\nu} \hat{h}_{\nu\mu}+\frac{1}{48} \hat{h}^3+\frac{\hat{r}^3}{9 \sqrt{6} A^6}\right)\right] +\mathcal{O}(\kappa^4) \ ,
\end{split}
\end{equation}
where $\hat{h}\equiv \eta^{\mu\nu}\hat{h}_{\mu\nu}$, and of the branes:
\begin{equation}
\label{detBR}
\begin{split}
   \sqrt{-g_\text{UV/IR}}=& A^4\left[ 1+\kappa  \left(\frac{1}{2} \hat{h}-\frac{\sqrt{\frac{2}{3}} \hat{r}}{A^2}\right)+\kappa ^2 \left(-\frac{\hat{r} \hat{h}}{\sqrt{6} A^2}-\frac{1}{4} \hat{h}^{\lambda\mu} \hat{h}_{\mu \lambda}+\frac{1}{8} \hat{h}^2+\frac{\hat{r}^2}{3 A^4}\right)+ \right.\\
   & \left. +\kappa ^3 \left(\frac{\hat{r} \hat{h}^{\lambda\mu} \hat{h}_{\mu\lambda}}{2 \sqrt{6} A^2}-\frac{\hat{r} \hat{h}^2}{4 \sqrt{6} A^2}+\frac{\hat{r}^2 \hat{h}}{6 A^4}+\frac{1}{6} \hat{h}^{\lambda\mu} \hat{h}_{\nu \lambda} \hat{h}_\mu^\nu -\frac{1}{8} \hat{h} \hat{h}^{\mu\nu} \hat{h}_{\nu\mu}+\frac{1}{48} \hat{h}^3-\frac{\sqrt{\frac{2}{3}} \hat{r}^3}{9 A^6}\right) \right]+ \\
   &+\mathcal{O}(\kappa^4) \ .
   \end{split}
\end{equation}
In the following we use a rescaled expansion parameter $\tilde{\kappa}\equiv \frac{1}{2}\kappa$. We are going to present the expansion of $\mathcal{L}_{\text{IR}}$ to $\mathcal{O}(\tilde{\kappa}^2)$ and of the RS-action up to $\mathcal{O}(\tilde{\kappa}^3)$. 
\subsection{$\mathbf{\mathcal{L}_{\text{IR}}}$-Expansion}
The IR-Lagrangian is given by 
\begin{equation}
    \mathcal{L}_\text{IR}=\sqrt{-g_\text{IR}}\left[\frac{1}{2}g_\text{IR}^{\mu\nu}\partial_\mu\phi \partial_\nu\phi - \frac{1}{2}m_\phi^2 \phi^2\right]
\end{equation}
We can expand the brane metric $g_\text{IR}$ and obtain the expansion
\begin{equation}
    \mathcal{L}_\text{IR}= \mathcal{L}^{(0)}_\text{IR}+\tilde{\kappa}\mathcal{L}^{(1)}_\text{IR}+\tilde{\kappa}^2\mathcal{L}^{(2)}_\text{IR}+\mathcal{O}(\tilde{\kappa}^3)
\end{equation}
{\bf First order:} The Lagrangian at first order is given by
\begin{equation}
     \mathcal{L}^{(1)}_\text{IR} =  \mathcal{L}^{(1)}_\text{IR}(\hat{h})+\mathcal{L}^{(1)}_\text{IR}(\hat{r}) \ ,
\end{equation}
with
\begin{equation}
    \begin{split}
        &\mathcal{L}^{(1)}_\text{IR}(\hat{h})= \frac{1}{2}  \left[- m_\phi^2  \hat{h} \phi ^2 + \phi {}^{,\mu  } ( \hat{h} \phi {}_{,\mu  } -2  \hat{h}_{\mu  \nu  } \phi {}^{,\nu})\right] \ , \\
        &\mathcal{L}^{(1)}_\text{IR}(\hat{r})= \frac{1}{\sqrt{6}} \hat{r} (2 m_\phi^2 \phi^2 - \phi {}_{,\mu  } \phi {}^{,\mu  }) \ .
    \end{split}
\end{equation}
 {\bf Second order:} At second order the expanded IR-Lagrangian is given by
\begin{equation}
     \mathcal{L}^{(2)}_\text{IR} =  \mathcal{L}^{(2)}_\text{IR}(\hat{h}^2)+\mathcal{L}^{(2)}_\text{IR}(\hat{h}\hat{r})+\mathcal{L}^{(2)}_\text{IR}(\hat{r}^2) \ ,
\end{equation}
with
\begin{equation}
  \begin{split}
        &\mathcal{L}^{(2)}_\text{IR}(\hat{h}^2)=\frac{1}{4} \left[m_\phi^2\phi^2 (2  \hat{h}^{\mu  \nu  } \hat{h}_{\nu  \mu  }  - \hat{h}^2 ) + \phi {}^{,\mu  } (-2 \hat{h}^{\nu  \lambda} \hat{h}_{\lambda\nu  } \phi{}_{,\mu  } + \hat{h}^2 \phi{}_{,\mu  } + 8 \hat{h}_{\mu  }{}^{\lambda} \hat{h}_{\lambda\nu  } \phi {}^{,\nu  } -4 \hat{h}_{\mu  \nu  } \hat{h} \phi {}^{,\nu})\right] \ ,\\
        &\mathcal{L}^{(2)}_\text{IR}(\hat{h}\hat{r})=\frac{1}{\sqrt{6}}\hat{r} (2 m_\phi^2 \hat{h} \phi ^2 + \phi {}^{,\mu  } (- \hat{h} \phi {}_{,\mu  } + 2 \hat{h}_{\mu \nu  } \phi {}^{,\nu})) \ ,\\
        &\mathcal{L}^{(2)}_\text{IR}(\hat{r}^2)=\frac{1}{6} \hat{r}^2 (-4 m_\phi^2 \phi^2 + \phi {}_{,\mu  } \phi {}^{,\mu  }) \ .
  \end{split}
\end{equation}
\subsection{Bulk-Expansion}
We expand the bulk Lagrangian as
\begin{equation}
    \mathcal{L} \subset \mathcal{L}_2+\tilde{\kappa} \ \mathcal{L}_3 +\mathcal{O}(\tilde{\kappa}^2) \ ,
\end{equation}
where $\mathcal{L}$ already includes the prefactors of the action and the determinant of the metric, i.e. $\mathcal{L}=\frac{1}{2}\sqrt{G}M^3_5 R$, such that
\begin{equation}
    S_{\text{RS}}=\int d^4x \int\limits_{-z_{\text{IR}}}^{z_{\text{IR}}} \diff{z} \left(\mathcal{L}-\sqrt{G}M^3_5\Lambda_B\right) \ .
\end{equation}
To make the discussion more clear and ordered, the Lagrangian will be separated in two parts depending on whether they contain  5D-derivatives or not; these will be denoted $\mathcal{L}^A$ and $\mathcal{L}^B$, respectively. Also some other terms are generated in the expansion; these are responsible for the vacuum energies cancellations and we do not report them here.
We separate each piece further depending on how many $\hat{h}$ and $\hat{r}$ fields they include.

{\bf Second order:}
$\mathcal{L}_2 = \mathcal{L}_2(\hat{h}^2)+\mathcal{L}_2(\hat{h}\hat{r})+\mathcal{L}_2(\hat{r}^2)$.\\
\begin{itemize}
    \item $\mathcal{L}_2(\hat{h}^2)$: 
\begin{equation}
    \begin{split}
        & \mathcal{L}_2^{A}(\hat{h}^2) =  \tfrac{1}{2} A^3 \left[\hat{h}\square\hat{h}-\hat{h}^{\mu\nu}\square\hat{h}_{\mu\nu}-2\hat{h}\hat{h}^{\mu\nu}{}_{,\mu\nu}+2\hat{h}^{\mu\nu}\hat{h}^{\rho}{}_{\nu,\rho\mu} \right] \ ,\\
        &\mathcal{L}_{2}^{B}(\hat{h}^2) =  -\frac{1}{2} A^2 \left[\hat{h} \left(3 \hat{h}' A'+A \hat{h}''\right)-\hat{h}^{\mu\nu}\left(3 \hat{h}_{\mu\nu}' A'+A \hat{h}_{\mu\nu}''\right)\right] \ .
    \end{split}
\end{equation}
\item $\mathcal{L}_2(\hat{h}\hat{r})$:
\begin{equation}
    \begin{split}
        & \mathcal{L}_2^{A}(\hat{h}\hat{r}) = 0\ ,\\
        &\mathcal{L}_{2}^{B}(\hat{h}\hat{r}) = 0 \ . 
    \end{split}
\end{equation}
\item $\mathcal{L}_2(\hat{r}^2)$:
\begin{equation}
    \begin{split}
        & \mathcal{L}_2^{A}(\hat{r}^2) = \frac{1}{2A} \ \hat{r}{}^{,\lambda }\hat{r}{}_{,\lambda }\ ,\\
        &\mathcal{L}_{2}^{B}(\hat{r}^2) =  0\ .
    \end{split}
\end{equation}
\end{itemize}
{\bf Third order:}
$\mathcal{L}_3 = \mathcal{L}_3(\hat{h}^3)+\mathcal{L}_3(\hat{h}^2\hat{r})+\mathcal{L}_3(\hat{h}\hat{r}^2)+\mathcal{L}_3(\hat{r}^3)$.

\begin{itemize}
    \item $\mathcal{L}_3(\hat{h}^3)$:
\begin{equation}
    \begin{split}
        & \mathcal{L}_3^{A}(\hat{h}^3) = A^3\left[\frac{1}{2}\hat{h}_{\mu\nu}\hat{h}_{\rho\sigma}\hat{h}^{\rho\sigma,\mu\nu}-\frac{1}{2}\hat{h}\hat{h}_{\mu\nu}\hat{h}^{,\mu\nu}-2\hat{h}_{\mu\nu}\hat{h}^{,\mu}\hat{h}^{\rho\nu}{}_{,\rho}-\hat{h}_{\mu\nu}\hat{h}^{\nu\rho}\hat{h}^{,\mu\sigma}{}_{\sigma\rho}+\hat{h}_{\mu\nu}\hat{h}_{\rho\sigma}{}^{,\mu}\hat{h}^{\nu\rho,\sigma}\right.\\
        &\left.\qquad \qquad -\frac{1}{4}\hat{h}\hat{h}^{\mu\nu}\square\hat{h}_{\mu\nu}+\frac{3}{4}\hat{h}_{\mu\nu}\hat{h}^{\mu\nu,\rho}\hat{h}_{,\rho}+\frac{1}{2}\hat{h}_{\mu\nu}\hat{h}^{\nu\rho}\square\hat{h}^{\mu}{}_{\rho}-\frac{1}{2}\hat{h}_{\mu\nu}\hat{h}_{\rho\sigma}{}^{,\rho}\hat{h}^{\mu\nu,\sigma}+\frac{1}{2}\hat{h}\hat{h}_{\nu\rho,\mu}\hat{h}^{\mu\rho,\nu}\right.\\
        &\left.\qquad \qquad -\hat{h}\hat{h}^{\mu\nu}{}_{,\mu}\hat{h}_{\nu\rho}{}^{,\rho}+\frac{1}{8}\hat{h}^2\square\hat{h}\right]\ ,\\
        &\mathcal{L}_{3}^{B}(\hat{h}^3) = A^2 \left[-\frac{1}{4} \hat{h}^2 \left(3\hat{h}'A'+A\hat{h}'' \right) + \frac{3}{4}\hat{h}_{\mu\nu}\hat{h}^{\mu\nu}\left(3\hat{h}'A'+A\hat{h}'' \right)+\frac{1}{2}\hat{h}^{\mu\nu}\hat{h}\left(3\hat{h}_{\mu\nu}'A'+A\hat{h}_{\mu\nu}'' \right) \right.\\
        & \left. \qquad \qquad -\hat{h}^{\mu\lambda}\hat{h}^\nu_\mu\left(3\hat{h}_{\nu\lambda}'A'+A\hat{h}_{\nu\lambda}'' \right)\right]\ .
    \end{split}
\end{equation}
\item $\mathcal{L}_3(\hat{h}^2\hat{r})$:
\begin{equation}
    \begin{split}
        & \mathcal{L}_3^{A}(\hat{h}^2\hat{r}) = 0 \ ,\\
        &\mathcal{L}_{3}^{B}(\hat{h}^2\hat{r}) = A^3\sqrt{\frac{3}{2}}\frac{\hat{r}}{A^2}\left[\hat{h}^{\mu\nu}{}'\hat{h}_{\mu\nu}'-(\hat{h}')^2 \right]\ . 
    \end{split}
\end{equation}
\item $\mathcal{L}_3(\hat{h}\hat{r}^2)$:
\begin{equation}
    \begin{split}
        & \mathcal{L}_3^{A}(\hat{h}\hat{r}^2) =\frac{1}{A}\left[ \hat{h}_{\mu\nu}\hat{r}\hat{r}^{,\mu\nu}-\frac{1}{6}\hat{h}_{\mu\nu}{}^{,\mu\nu}\hat{r}^2 -\frac{1}{2}\hat{h}\hat{r}\Box\hat{r}-\frac{1}{12}\Box\hat{h}\hat{r}^2\right]\ ,\\
        &\mathcal{L}_{3}^{B}(\hat{h}\hat{r}^2) = 0 \ .
    \end{split}
\end{equation}
\item $\mathcal{L}_3(\hat{r}^3)$:
\begin{equation}
    \begin{split}
        & \mathcal{L}_3^{A}(\hat{r}^3) = -\sqrt{\frac{2}{3}}\frac{\hat{r}\ \hat{r}^{,\mu}\hat{r}_{,\mu}}{A^3}\ ,\\
        &\mathcal{L}_{3}^{B} (\hat{r}^3)=  0\ .
    \end{split}
\end{equation}
\end{itemize}

\section{Feynman Rules \label{Appx:Feynman}}
List of the Feynman rules that are relevant for the considered processes. At the vertex all momenta are taken to be directed inwards.
\paragraph{Propagators:}$\,$\newline
\begin{minipage}[h]{.45\textwidth}
 \begin{figure}[H]
      \includegraphics[scale=0.7]{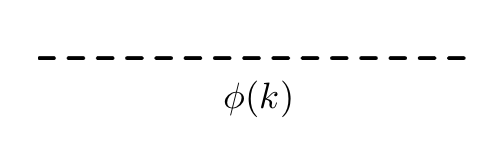}
\end{figure}
\end{minipage}\hfill
\begin{minipage}[h]{.55\textwidth}
\begin{flalign}
   = \frac{i}{k^2 - m_\phi^2} \ ,&&
    \end{flalign}
\end{minipage}
\begin{minipage}[h]{.45\textwidth}
 \begin{figure}[H]
      \includegraphics[scale=0.7]{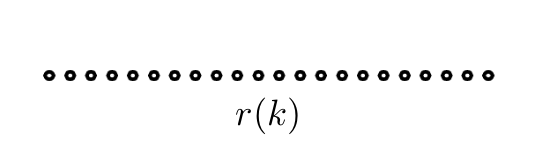}
\end{figure}
\end{minipage}\hfill
\begin{minipage}[h]{.55\textwidth}
\begin{flalign}
   =  \frac{i}{k^2 - m_r^2 } \ ,&&
    \end{flalign}
\end{minipage}
\begin{minipage}[h]{.45\textwidth}
 \begin{figure}[H]
      \includegraphics[scale=0.7]{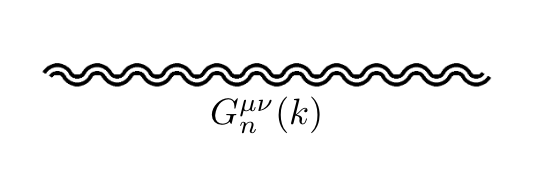}
\end{figure}
\end{minipage}\hfill
\begin{minipage}[h]{.55\textwidth}
\begin{flalign}
   =  \frac{i P_n^{\mu\nu\alpha\beta}(k)}{k^2 - m_n^2 } \ ,&&
    \end{flalign}
\end{minipage}
where 
\begin{equation}
    \begin{split}
        P^n_{\mu\nu\alpha\beta}(p) \equiv \sum\limits_s  \epsilon^s_{\mu\nu}(p)\epsilon^s_{\alpha\beta}(p)^* &= \frac{1}{2}\left(G_{\mu\alpha}G_{\nu\beta}+G_{\nu\alpha}G_{\mu\beta}-\frac{2}{3}G_{\mu\nu}G_{\alpha\beta} \right) \ ,\\
        \text{with} &\quad G_{\mu\nu}\equiv \eta_{\mu\nu}-\frac{p_\mu p_\nu}{m_n^2}\ .
    \end{split}
\end{equation}

\subsection{Vertices}
For convenience, we define the quantity:
\begin{equation}
    C_{\mu\nu\alpha\beta}\equiv \eta_{\mu\alpha}\eta_{\nu\beta}+\eta_{\mu\beta}\eta_{\nu\alpha}-\eta_{\mu\nu}\eta_{\alpha\beta} \ .
\end{equation}
With this:
\begin{equation}
   \begin{split}
       \mathcal{A}^{\mu\nu\alpha\beta\iota\kappa}\equiv \ & \eta^{\alpha  \iota } C^{\beta  \kappa\mu  \nu }+\eta^{\alpha  \kappa } C^{\beta  \iota \mu  \nu}+\eta^{\alpha  \mu } C^{\beta  \nu \iota  \kappa}+\eta^{\alpha  \nu } C^{\beta  \mu \iota  \kappa } -2\eta^{\mu\nu}C^{\alpha\beta\iota\kappa}-\eta^{\alpha  \beta } \left(\eta^{\iota  \nu } \eta^{\kappa  \mu }+ \eta^{\iota  \mu } \eta^{\kappa  \nu }\right) \ ,
   \end{split}
\end{equation}
\begin{equation}
\begin{split}
    \mathcal{B}^{\mu\nu\alpha\beta\iota\kappa}[k]\equiv \ &-C^{\iota  \kappa \alpha  \beta }k^{\mu } k^{\nu } -2C^{\mu\nu\alpha\beta} k^{\iota }k^{\kappa}-2C^{\mu\nu\iota\kappa} k^{\alpha}k^{\beta} \\
    &+\left(- \eta^{\beta  \kappa } \eta^{\mu  \nu } k^{\alpha }- \eta^{\alpha  \kappa } \eta^{\mu  \nu } k^{\beta } +C^{\beta\kappa\mu\nu}k^{\alpha}+C^{\alpha\kappa\mu\nu}k^{\beta}\right) k^{\iota }\\
    &+\left(- \eta^{\beta  \iota } \eta^{\mu  \nu } k^{\alpha }- \eta^{\alpha  \iota } \eta^{\mu  \nu } k^{\beta } +C^{\beta\iota\mu\nu}k^{\alpha}+C^{\alpha\iota\mu\nu}k^{\beta}\right) k^{\kappa }\\
    &+\left(C^{\beta  \mu\iota  \kappa} k^{\alpha }+C^{\alpha  \mu \iota  \kappa} k^{\beta }+C^{\alpha  \beta\kappa  \mu } k^{\iota }+C^{\iota  \mu \alpha  \beta } k^{\kappa }\right) k^{\nu }\\
    &+\left(C^{\beta  \nu\iota  \kappa} k^{\alpha }+C^{\alpha  \nu \iota  \kappa} k^{\beta }+C^{\alpha  \beta\kappa  \nu } k^{\iota }+C^{\iota  \nu \alpha  \beta } k^{\kappa }\right) k^{\mu }\\
    & -\left(\mathcal{A}^{\mu\nu\alpha\beta\iota\kappa}+2\eta^{\mu\nu}C^{\alpha\beta\iota\kappa}+\eta^{\alpha  \beta } \eta^{\iota  \kappa } \eta^{\mu  \nu }\right) k^2 \ ,
\end{split}
\end{equation}
\begin{equation}
\begin{split}
    \mathcal{C}^{\mu\nu\alpha\beta\iota\kappa}[k,p]\equiv & \ \left(C^{\beta  \kappa \mu  \nu } p^{\iota }+C^{\beta  \iota \mu  \nu }p^{\kappa }+C^{\beta  \nu\iota  \kappa} p^{\mu }+C^{\beta\mu\iota\kappa} p^{\nu } -2C^{\iota\kappa\mu\nu} p^{\beta } - p^{\kappa } \eta^{\beta  \iota } \eta^{\mu  \nu } - p^{\iota } \eta^{\beta  \kappa } \eta^{\mu  \nu }\right) k^{\alpha }\\
    &+\left(C^{\alpha  \kappa \mu  \nu } p^{\iota }+C^{\alpha  \iota \mu  \nu }p^{\kappa }+C^{\alpha  \nu\iota  \kappa} p^{\mu }+C^{\alpha\mu\iota\kappa} p^{\nu } -2C^{\iota\kappa\mu\nu} p^{\alpha } - p^{\kappa } \eta^{\alpha  \iota } \eta^{\mu  \nu } - p^{\iota } \eta^{\alpha  \kappa } \eta^{\mu  \nu } \right)k^{\beta }\\
    &+\left(C^{\alpha  \beta \kappa  \nu } p^{\mu } +C^{\alpha  \beta \kappa  \mu } p^{\nu } +2 C^{\mu  \nu\beta  \kappa} p^{\alpha } +2 C^{\alpha  \kappa\mu  \nu } p^{\beta } -3 C^{\alpha  \beta\mu  \nu }p^{\kappa } - \eta^{\alpha  \beta } \eta^{\kappa  \nu }p^{\mu }\right) k^{\iota }\\
    &+\left(C^{\alpha  \beta \iota  \nu } p^{\mu } +C^{\alpha  \beta \iota  \mu } p^{\nu } +2 C^{\mu  \nu\beta  \iota} p^{\alpha } +2 C^{\alpha  \iota\mu  \nu } p^{\beta } -3 C^{\alpha  \beta\mu  \nu }p^{\iota } - \eta^{\alpha  \beta } \eta^{\iota  \nu }p^{\mu }\right)k^{\kappa }\\
    &-\eta^{\alpha  \beta }\left(\left(\eta^{\kappa  \mu }p^{\nu } +  \eta^{\mu  \nu } p^{\kappa } \right) k^{\iota }+\left( \eta^{\iota  \mu }p^{\nu } +  \eta^{\mu  \nu } p^{\iota }  \right)k^{\kappa } \right)\\
    &-2 \left(-C^{\iota\kappa\beta\nu} p^{\alpha }-C^{\iota\kappa\alpha\nu}p^{\beta }-C^{\alpha\beta\kappa\nu} p^{\iota }-C^{\iota  \nu \alpha  \beta } p^{\kappa }+C^{\iota  \kappa \alpha  \beta } p^{\nu }\right) k^{\mu }\\
    &-2 \left(-C^{\iota\kappa\beta\mu} p^{\alpha }-C^{\iota  \kappa \alpha  \mu } p^{\beta }-C^{\alpha  \beta \kappa  \mu } p^{\iota }-C^{\iota  \mu \alpha  \beta } p^{\kappa } +C^{\iota  \kappa \alpha  \beta } p^{\mu }\right) k^{\nu }\\
    &-\left(k \cdot  p\right)\left(3\mathcal{A}^{\mu\nu\alpha\beta\iota\kappa} +7\eta^{\mu\nu}C^{\alpha\beta\iota\kappa}+3\eta^{\mu\nu}\eta^{\alpha\beta}\eta^{\iota\kappa}+\eta^{\alpha\beta}\left(\eta^{\kappa\mu}\eta^{\iota\nu}+\eta^{\iota\mu}\eta^{\kappa\nu}\right)\right) \ .
\end{split}
\end{equation}
\subsubsection{Vertices involving only RS particles}
\begin{minipage}[h]{.25\textwidth}
 \begin{figure}[H]
      \includegraphics[scale=0.3]{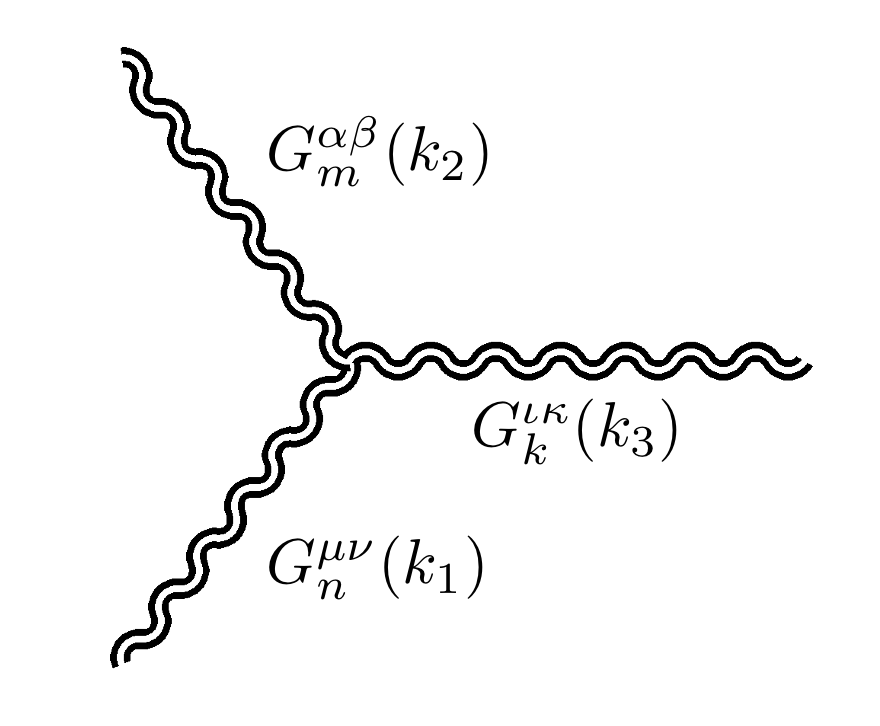}
\end{figure}
\end{minipage}\hfill
\begin{minipage}[h]{.75\textwidth}
\begin{flalign}
   = i\frac{\chi_{\text{nmk}}}{4\Lambda}\left[\mathcal{A}^{\mu\nu\alpha\beta\iota\kappa}m_n^2 + \mathcal{A}^{\alpha\beta\mu\nu\iota\kappa}m_m^2+\mathcal{A}^{\iota\kappa\mu\nu\alpha\beta}m_k^2  \right. && \\
    \nonumber \left. + 4\left(\mathcal{B}^{\mu\nu\alpha\beta\iota\kappa}[k_1] + \mathcal{B}^{\alpha\beta\mu\nu\iota\kappa}[k_2] + \mathcal{B}^{\iota\kappa\mu\nu\alpha\beta}[k_3]\right)\right. && \\
     \nonumber \left.+2\left(\mathcal{C}^{\mu\nu\alpha\beta\iota\kappa}[k_1,k_2]+\mathcal{C}^{\mu\nu\iota\kappa\alpha\beta}[k_1,k_3]+\mathcal{C}^{\alpha\beta\iota\kappa\mu\nu}[k_2,k_3]\right) \right] \ . &&
    \end{flalign}
\end{minipage}
\begin{minipage}[h]{.35\textwidth}
 \begin{figure}[H]
       \includegraphics[scale=0.35]{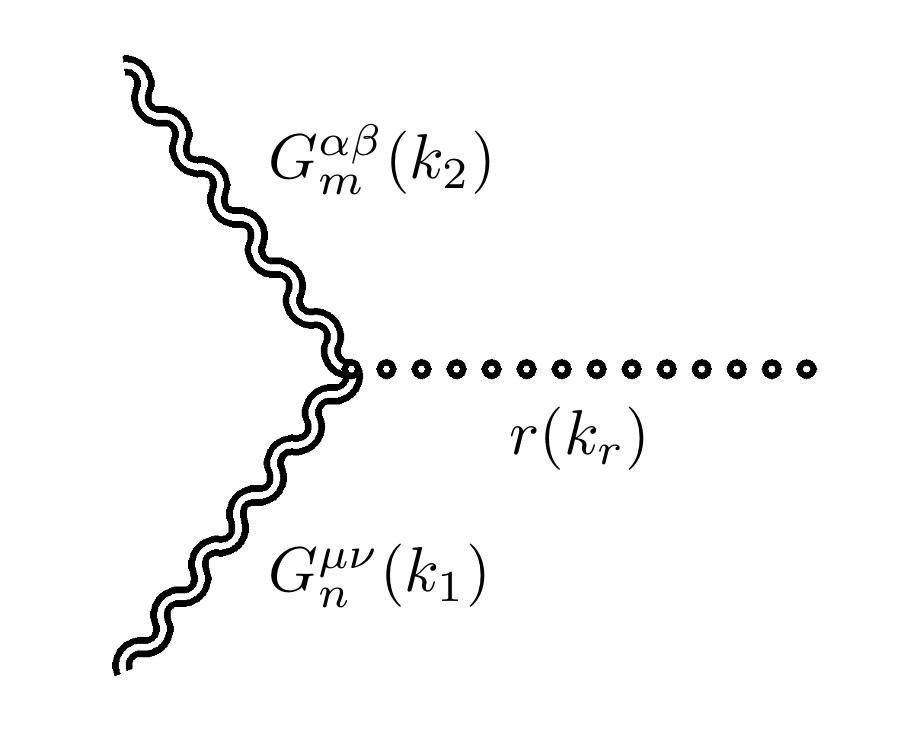}
\end{figure}
\end{minipage}\hfill
\begin{minipage}[h]{.65\textwidth}
\begin{flalign}
   = \frac{i \sqrt{\frac{3}{2}} e^{-\pi  \mu } \mu  \tilde{\chi} _{\text{nkr}} \left(-2 \eta^{\alpha  \beta } \eta^{\mu  \nu }+\eta^{\alpha  \mu } \eta^{\beta  \nu }+\eta^{\alpha  \nu } \eta^{\beta  \mu }\right)}{\Lambda  r_c} \ .&&
    \end{flalign}
\end{minipage}
\begin{minipage}[h]{.35\textwidth}
 \begin{figure}[H]
       \includegraphics[scale=0.35]{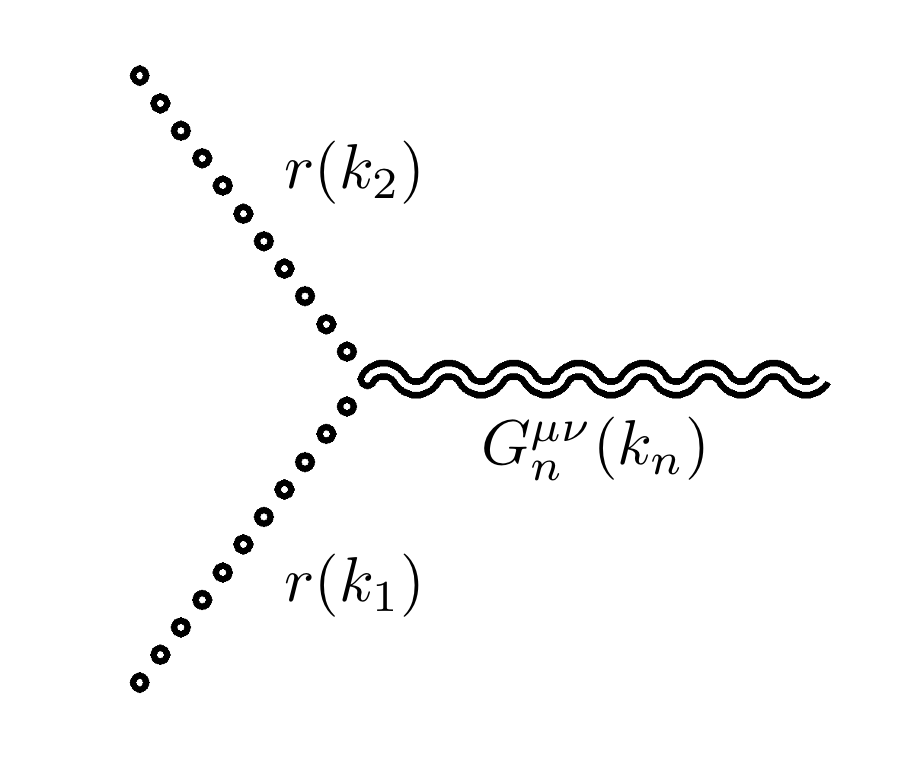}
\end{figure}
\end{minipage}\hfill
\begin{minipage}[h]{.65\textwidth}
\begin{flalign}
   = \frac{i \chi _{\text{nrr}}}{3 \Lambda } \left(2 (k_1^2+ k_2^2 ) \eta^{\mu  \nu }+\left(  k_1 \cdot k_2\right) \eta^{\mu  \nu }+k_1^{\mu } \left(k_2^{\nu }-2 k_1^{\nu }\right) \right. && \\
  \nonumber  \left. + k_2^{\mu }( k_1^{\nu }-2 k_2^{\nu })\right)\ . &&
    \end{flalign}
\end{minipage}
\begin{minipage}[h]{.35\textwidth}
 \begin{figure}[H]
       \includegraphics[scale=0.35]{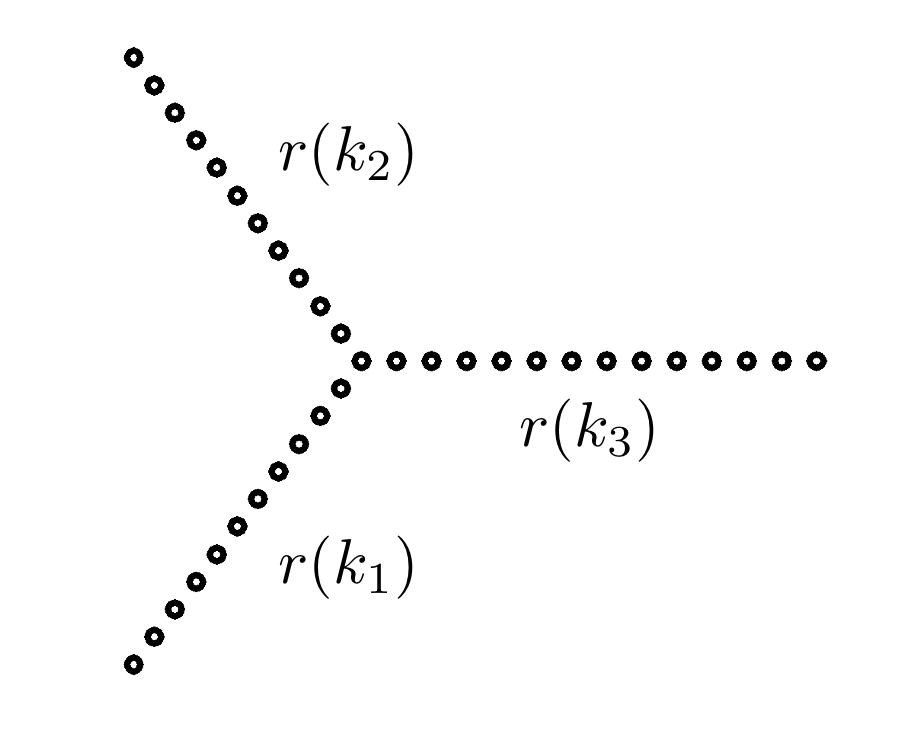}
\end{figure}
\end{minipage}\hfill
\begin{minipage}[h]{.65\textwidth}
\begin{flalign}
   = -\frac{i \sqrt{\frac{2}{3}} \left(k_1^2+k_2^2+k_3^2\right) \chi _{\text{rrr}}}{\Lambda }\ . &&
    \end{flalign}
\end{minipage}
\subsubsection{Vertices involving $\phi$}
\begin{minipage}[h]{.35\textwidth}
 \begin{figure}[H]
       \includegraphics[scale=0.35]{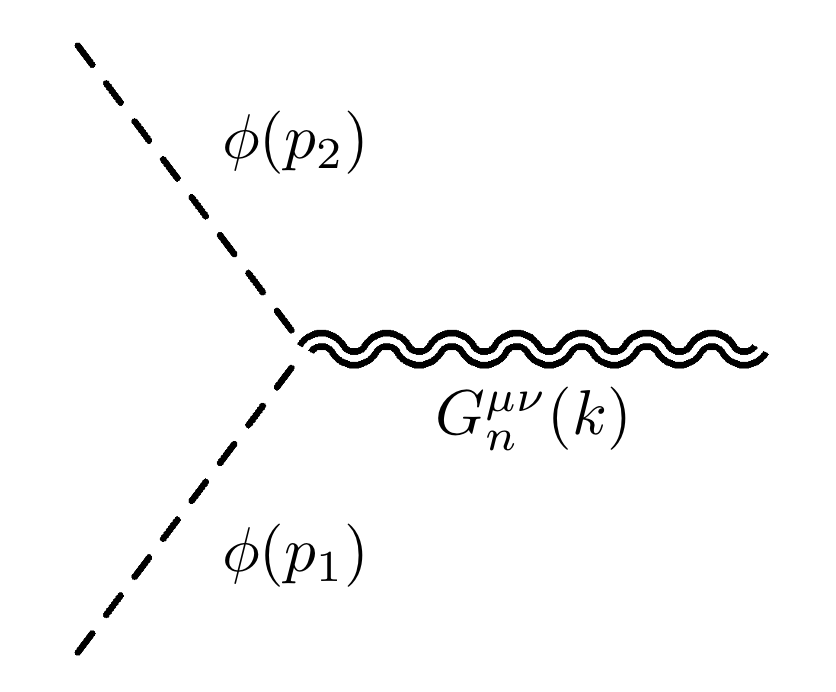}
\end{figure}
\end{minipage}\hfill
\begin{minipage}[h]{.65\textwidth}
\begin{flalign}
   = \frac{i \left(C^{\mu\nu\alpha\beta}{\text{p}_1}_{\alpha } {\text{p}_2}_{\beta }-{\eta}^{\mu  \nu } m_{\phi }^2\right)}{\Lambda }\ . &&
    \end{flalign}
\end{minipage}
\begin{minipage}[h]{.35\textwidth}
 \begin{figure}[H]
        \includegraphics[scale=0.35]{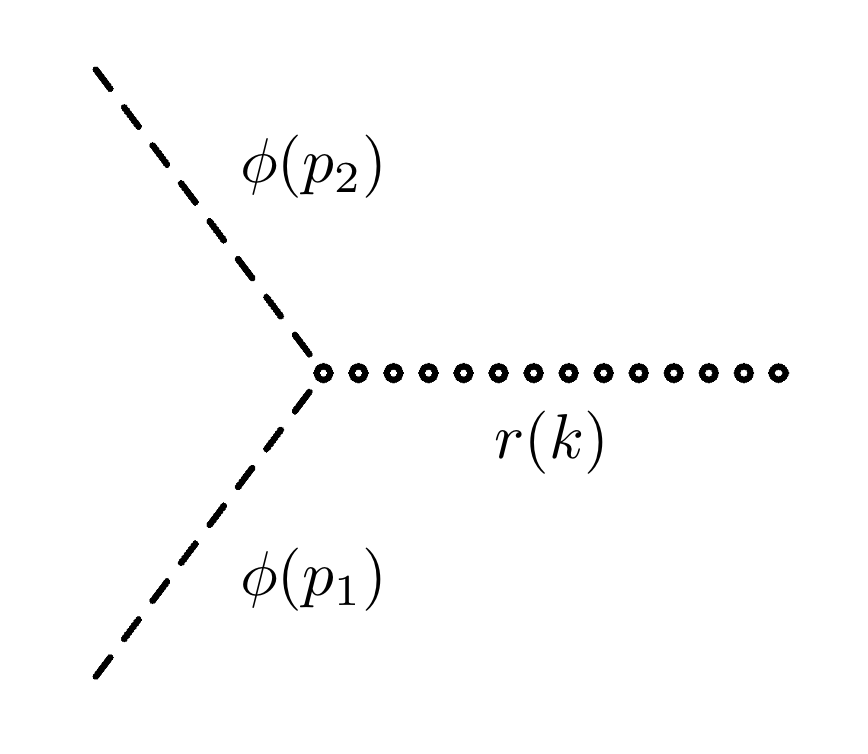}
\end{figure}
\end{minipage}\hfill
\begin{minipage}[h]{.65\textwidth}
\begin{flalign}
  = \frac{i \sqrt{\frac{2}{3}} \left( {p}_1\cdot  {p}_2+2 m_{\phi }^2\right)}{\Lambda }\ . &&
    \end{flalign}
\end{minipage}

\begin{minipage}[h]{.35\textwidth}
 \begin{figure}[H]
         \includegraphics[scale=0.35]{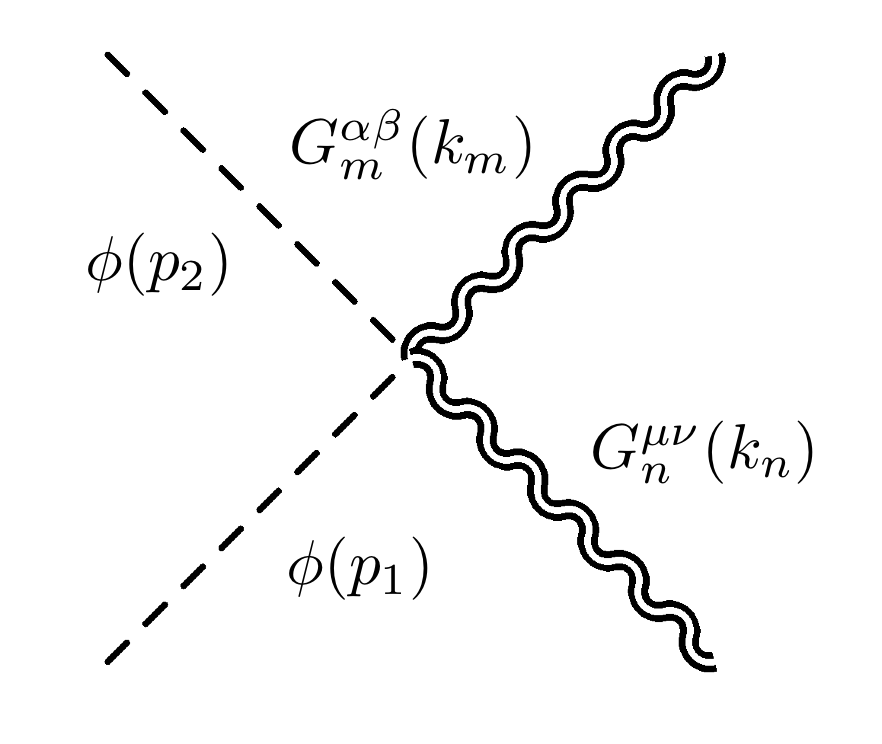}
\end{figure}
\end{minipage}\hfill
\begin{minipage}[h]{.65\textwidth}
\begin{flalign}
   = -\frac{i}{\Lambda ^2} \left(-m_{\phi }^2 C^{\mu\nu\alpha\beta}+p_1^{\alpha } \left(-\eta^{\mu  \nu }p_2^{\beta } +p_2^{\mu } \eta^{\beta  \nu }+p_2^{\nu } \eta^{\beta  \mu }\right)\right. && \\
  \nonumber \left. +p_1^{\beta } \left(-\eta^{\mu  \nu }p_2^{\alpha } +p_2^{\mu } \eta^{\alpha  \nu }+p_2^{\nu } \eta^{\alpha  \mu }\right)+p_2^{\alpha } p_1^{\mu } \eta^{\beta  \nu }+p_2^{\beta } p_1^{\mu } \eta^{\alpha  \nu } \right. && \\
   \nonumber\left. -p_1^{\mu } p_2^{\nu } \eta^{\alpha  \beta }+p_2^{\alpha } p_1^{\nu } \eta^{\beta  \mu }+p_2^{\beta } p_1^{\nu } \eta^{\alpha  \mu }-p_2^{\mu } p_1^{\nu } \eta^{\alpha  \beta } \right. && \\
   \nonumber \left. -\left(  p_1 \cdot p_2\right) \eta^{\alpha  \nu } \eta^{\beta  \mu }-\left(  p_1\cdot p_2\right) \eta^{\alpha  \mu } \eta^{\beta  \nu }+\left(  p_1\cdot p_2\right) \eta^{\alpha  \beta } \eta^{\mu  \nu }\right) \ .&&
    \end{flalign}
\end{minipage}
\begin{minipage}[h]{.35\textwidth}
 \begin{figure}[H]
          \includegraphics[scale=0.35]{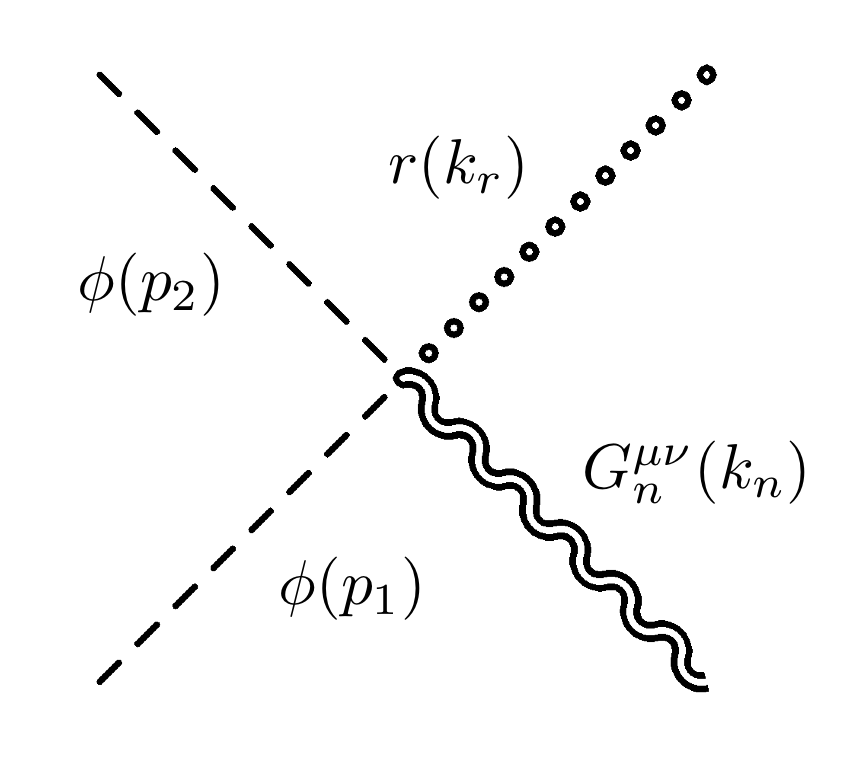}
\end{figure}
\end{minipage}\hfill
\begin{minipage}[h]{.65\textwidth}
\begin{flalign}
    = -\frac{i \sqrt{\frac{2}{3}} \left(-\eta^{\mu  \nu } \left(  p_1\cdot p_2+2 m_{\phi }^2\right)+p_1^{\mu } p_2^{\nu }+p_2^{\mu } p_1^{\nu }\right)}{\Lambda ^2} \ .&&
    \end{flalign}
\end{minipage}
\begin{minipage}[h]{.35\textwidth}
 \begin{figure}[H]
          \includegraphics[scale=0.35]{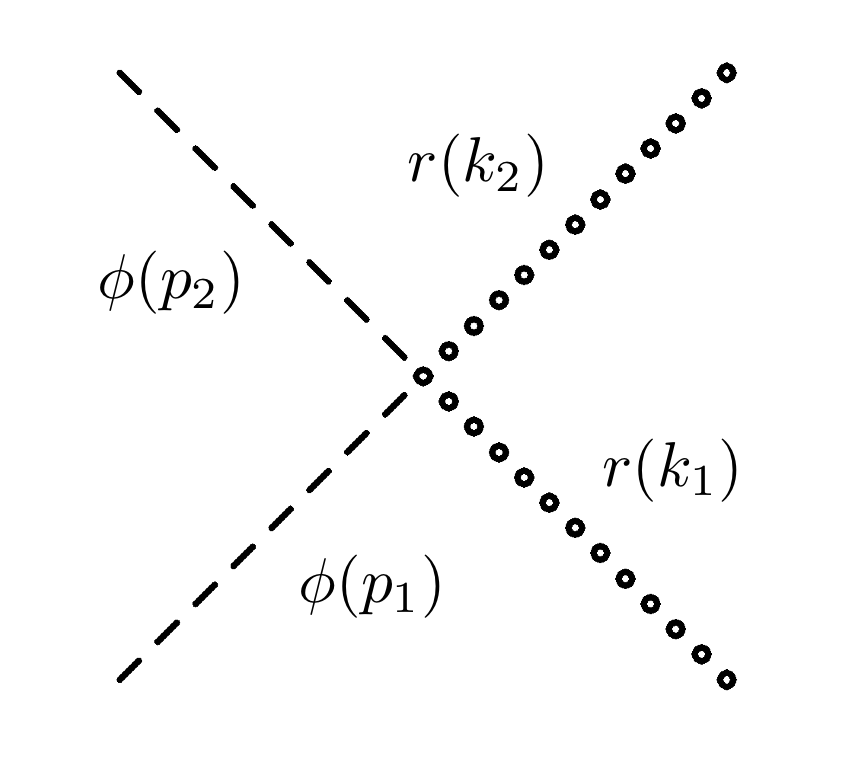}
\end{figure}
\end{minipage}\hfill
\begin{minipage}[h]{.65\textwidth}
\begin{flalign}
    = -\frac{2 i \left( p_1 \cdot p_2+4 m_{\phi }^2\right)}{3 \Lambda ^2}\ . &&
    \end{flalign}
\end{minipage}

\end{appendices}

\bibliographystyle{hieeetr}
\bibliography{extra_dim.bib}

\end{document}